\documentclass[preprint,tightenlines,aps,prd,showpacs,nofootinbib]{revtex4}
\usepackage{graphicx}
\usepackage{amsmath}
\usepackage{amssymb}
\usepackage{bm}

\begin{document}
\preprint{MIT-CTP 4454}
\title{\mbox{}\\[10pt]
The $\bm{J/\psi\, \omega}$ Decay Channel\\
 of the $\bm{X(3872)}$ Charm Meson Molecule}

\author{Eric Braaten}
\affiliation{Physics Department, Ohio State University, Columbus,
Ohio 43210, USA}

\author{Daekyoung Kang}
\affiliation{Center for Theoretical Physics, 
Massachusetts Institute of Technology, 
Cambridge, MA 02139, USA}
\date{\today}
\begin{abstract}
Analyses of the $J/\psi\, \pi^+ \pi^-$ decay channel 
of the $X(3872)$ resonance by the CDF, Belle, and LHCb collaborations 
have established its $J^{PC}$ quantum numbers as $1^{++}$.
An analysis of the $\pi^+ \pi^- \pi^0$ invariant mass distribution
in the $J/\psi\,  \pi^+ \pi^- \pi^0$ decay channel 
by the Babar collaboration indicated a preference for $2^{-+}$ over $1^{++}$.
We point out that  a proper evaluation of the $\chi^2$
in that analysis increases the probability for $1^{++}$ 
from 7.1\% to about 18.7\%.
In the case of quantum numbers $1^{++}$, 
where the $X$ has an S-wave coupling to $J/\psi\, \omega$,
the proximity of the $J/\psi\, \omega$ threshold to $D^* \bar D$ thresholds 
and the narrow width of the $\omega$ 
suggest that the effects of scattering between $J/\psi\, \omega$ 
and charm meson pairs could be significant.
We derive invariant mass distributions for $J/\psi\, \pi^+ \pi^- \pi^0$
and $\pi^+ \pi^- \pi^0$ that take into account S-wave scattering between the 
$D^{*0} \bar D^0$, $D^{*+} D^-$, and $J/\psi\, \omega$ channels.  
We also analyze the effects of scattering through the  
$\chi_{c1}(2P)$ charmonium resonance.
We find that scattering effects are unable to produce
significant changes in the shape of the  $\pi^+ \pi^- \pi^0$ invariant mass distribution.
\end{abstract}



\maketitle


\section{Introduction}

Ever since the discovery of the $X(3872)$ by the Belle Collaboration 
in 2003 \cite{Choi:2003ue}, one of the leading interpretations 
has been a charm meson molecule whose constituents are a superposition of 
$D^{*0} \bar D^0$ and $D^0 \bar D^{*0}$ \cite{Close:2003sg,Voloshin:2003nt}.
This identification would require the $J^{PC}$ quantum numbers of $X(3872)$ to be $1^{++}$.
The observation of its decay into $J/\psi\, \gamma$
determined the charge conjugation $C$ to be $+$ \cite{Abe:2005ix,Aubert:2006aj}.
In 2006, the CDF Collaboration reduced the options for $J^{PC}$ to 
$1^{++}$ and $2^{-+}$ by  analyzing decays into $J/\psi\, \pi^+ \pi^-$ \cite{Abulencia:2006ma}.
The LHCb Collaboration recently ruled out $2^{-+}$,  
finally establishing the quantum numbers of $X(3872)$ as $1^{++}$ \cite{Aaij:2013zoa}.

The option $2^{-+}$ had been disfavored on various theoretical grounds, 
especially if the $X(3872)$ is identified with the $^1D_2$ charmonium state $\eta_{c2}'$.
The prediction of the mass of $\eta_{c2}'$ in most potentials
models is  lower than 3872~MeV by 40 to 100~MeV \cite{Barnes:2003vb}.
The decay of $\eta_{c2}'$ into $J/\psi\, \gamma$ should 
have strong multipole suppression \cite{Jia:2010jn}.
The expected production rate for D-wave charmonium in a hadron collider
is much smaller than the observed production rate of the $X(3872)$ at the 
Tevatron \cite{Burns:2010qq}.  The decay of D-wave charmonium  into $D^0 \bar D^0 \pi^0$
should have angular momentum suppression \cite{Kalashnikova:2010hv}.
Finally, the degree of isospin violation required 
by the observed branching fraction into $J/\psi\, \pi^+ \pi^-$  is difficult 
to accommodate for D-wave charmonium \cite{Hanhart:2011tn}.
All of these problems are solved, or at least ameliorated, 
if the quantum numbers are $1^{++}$.

Back in 2010, the Babar collaboration 
analyzed decays of  $X(3872)$ into $J/\psi\, \pi^+ \pi^- \pi^0$ 
and concluded that $2^{-+}$ was preferred over 
$1^{++}$ \cite{delAmoSanchez:2010jr}.
They quantified this preference in terms of a probability that was 
7.1\% for $1^{++}$ and 61.9\% for $2^{-+}$.
We will point out that a proper quantification 
of the likelihood for the observed result increases 
the probability for $1^{++}$ to 18.7\%.
With the properly calculated probabilities, the preference for 
$2^{-+}$ over $1^{++}$ is no longer so significant.
However it is still worth considering whether a more accurate 
description of the resonance in the $J/\psi\, \pi^+ \pi^- \pi^0$ channel
would be important in the Babar analysis or in future analyses.

Since it has quantum numbers $1^{++}$, 
the $X$ has an S-wave coupling to $J/\psi\, \omega$.
The proximity of the $J/\psi\, \omega$ threshold to 
the $D^{*0} \bar D^0$ and $D^{*+} D^-$ thresholds 
and the narrow width of the $\omega$ 
suggest that the effects of scattering between $J/\psi\, \omega$ 
and charm meson pairs could be significant.
We therefore study the effects of scattering between these coupled channels
on the $X(3872)$ resonance in the  $J/\psi\, \pi^+ \pi^- \pi^0$ channel.
We also analyze the effects of scattering through the  
$\chi_{c1}(2P)$ charmonium resonance, which has quantum numbers $1^{++}$.

In Section~\ref{sec:widths},
we introduce our notation for the three coupled channels 
and for the many masses that are relevant to this problem.
In Section~\ref{sec:DDscat}, we derive the scattering amplitudes 
due to S-wave scattering between the coupled channels.
We use them in Section~\ref{sec:lineshape} 
to determine the inclusive line shape of the $X(3872)$ resonance
and its line shape in the $J/\psi\, \pi^+ \pi^- \pi^0$ channel.
We also determine the effect of the $\chi_{c1}(2P)$ resonance 
on the line shape.
In Section~\ref{sec:3pionmass}, we derive a simple expression
for the $\pi^+ \pi^- \pi^0$ invariant mass distribution.  
We examine the Babar results in Ref.~\cite{delAmoSanchez:2010jr}
and point out that a proper evaluation of the $\chi^2$ 
significantly increases the probability for the quantum numbers $1^{++}$.  
We show that the experimental resolution,
which was ignored in previous theoretical analyses, has a significant effect 
on the $\pi^+ \pi^- \pi^0$ invariant mass distribution.
Finally we study the effect on that invariant mass distribution 
of scattering between the three coupled channels and scattering through the  
$\chi_{c1}(2P)$ charmonium resonance.

\section{Notation and masses}
\label{sec:widths}

We consider the effects of scattering between three 
$J^{PC} = 1^{++}$ channels involving the particle pairs
$D^{*0} \bar D^0$, $D^{*+} D^-$, and $J/\psi\, \omega$.
We label the three channels by the integers 0, 1, and 2
and a vector index $n$ associated with the polarizations 
of the spin-1 particles:
\begin{subequations}
\begin{eqnarray}
| 0, n \rangle &=&  
\frac{1}{\sqrt2}\Big( | D^{*0}(n)\, \bar D^0 \rangle - | D^0\, \bar D^{*0}(n) \rangle \Big),
\label{ch0}
\\
| 1, n \rangle &=&  
\frac{-1}{\sqrt2}\Big( | D^{*+}(n)\, D^- \rangle - | D^+\, D^{*-}(n) \rangle \Big),
\label{ch1}
\\
| 2, n \rangle &=&  
\frac{\epsilon_{n m l}}{\sqrt2} \, | J/\psi(m)\, \omega(l) \rangle.
\label{ch2}
\end{eqnarray}
\label{ch012}
\end{subequations}
The $3 \times 3$ matrices that project these channels onto 
isospin 0 and isospin 1 are 
\begin{subequations}
\begin{eqnarray}
\Pi_0 &=&  
\left( 
\begin{array}{c c c}
~~\mbox{$\frac12$} & - \mbox{$\frac12$} & ~~0~~ \\
- \mbox{$\frac12$} & ~~\mbox{$\frac12$} & ~~0~~ \\
         ~~0     &           ~~0    & ~~1~~ 
\end{array} \right),
\label{ispin0}
\\
\Pi_1 &=&  
\left( 
\begin{array}{c c c}
~~\mbox{$\frac12$} & ~~\mbox{$\frac12$} & ~~0~~ \\
~~\mbox{$\frac12$} & ~~\mbox{$\frac12$} & ~~0~~ \\
         ~~0     &           ~~0    & ~~0~~ 
\end{array} \right).
\label{ispin1}
\end{eqnarray}
\label{ispin01}
\end{subequations}

We denote the masses of the charm mesons $D^{*0}$, $D^0$, $D^{*+}$, 
and $D^+$ by $M_{*0}$, $M_0$, $M_{*1}$, and $M_1$
and the masses of $J/\psi$ and $\omega$ by $M_\psi$ and $M_\omega$.
We denote the reduced masses for the three channels in Eqs.~(\ref{ch012}) 
by $\mu_0$, $\mu_1$, and $\mu_{\psi \omega}$, respectively.
The energy differences $\delta_1$ and $\delta_{\psi \omega}$
between the thresholds for $D^{*+} D^-$ and $J/\psi\, \omega$
and the $D^{*0} \bar D^0$ threshold are
\begin{subequations}
\begin{eqnarray}
\delta_1 &=&  
(M_{*1} + M_1) - (M_{*0} + M_0)  \approx   8.1~{\rm MeV},
\label{delta1}
\\
\delta_{\psi \omega} &=&  
(M_\psi + M_\omega) - (M_{*0} + M_0) \approx  7.7~{\rm MeV}.
\label{deltaV}
\end{eqnarray}
\label{delta012}
\end{subequations}
We denote the total energy of the pair of particles in their center-of-momentum frame 
by $M$.  Their total energy relative to the $D^{*0} \bar D^0$ threshold is
\begin{equation}
E = M - (M_{*0} + M_0).
\label{energy}
\end{equation}
The amplitude for the propagation of a pair of particles
between contact interactions involves the square root of their total energy 
relative to threshold.
The appropriate thresholds for the pairs of particles in the channels 
in Eqs.~(\ref{ch012}) are complex,
with imaginary parts given by the sum of the decay widths of the two particles.
If one of the widths is much larger than the other one, 
it is sufficient to only take the larger one into account.
The resulting threshold factors
for the pairs of particles in the channels in Eqs.~(\ref{ch012}) are
\begin{subequations}
\begin{eqnarray}
\kappa(E) &=&  
\left[ -2 \mu_0 (E +  i \Gamma_{*0}/2) \right]^{1/2} ,
\label{kappa0}
\\
\kappa_1(E) &=&  
\left[ -2 \mu_1 (E - \delta_1 +  i \Gamma_{*1}/2) \right]^{1/2} ,
\label{kappa1}
\\
\kappa_{\psi \omega}(E) &=&  
\left[ -2 \mu_{\psi \omega} (E - \delta_{\psi \omega} + i \Gamma_\omega/2) \right]^{1/2} ,
\label{kappaV}
\end{eqnarray}
\label{kappa012}
\end{subequations}
where $\Gamma_{*0} \approx 66$~keV,  $\Gamma_{*1}\approx 96$~keV, 
and $\Gamma_\omega \approx 8.5$~MeV
are the decay widths of $D^{*0}$, $D^{*+}$, and $\omega$.
The reduced masses are $\mu_0=966.7$~MeV, $\mu_1=968.7$~MeV, and $\mu_{\psi\omega}=624.8$~MeV.
It is convenient to introduce a $3 \times 3$ matrix $K(E)$
whose diagonal entries are the threshold factors
in Eqs.~(\ref{kappa012}):
\begin{eqnarray}
K(E) =  
\left( 
\begin{array}{c c c}
\kappa(E) &     0     &     0      \\
    0   & \kappa_1(E) &     0      \\
    0   &     0      & \kappa_{\psi \omega}(E) 
\end{array} \right).
\label{Kmatrix}
\end{eqnarray}

We denote the mass of the $X(3872)$ by $M_X$ and its width by $\Gamma_X$.
The most precise determinations of $M_X$ and $\Gamma_X$
come from the $J/\psi\, \pi^+ \pi^-$ decay channel.
Measurements in this channel avoid biases associated with
the $D^{*0} \bar D^0$ threshold that plague some other 
decay channels, such as $D^0 \bar D^0 \pi^0$ \cite{Stapleton:2009ey}.
The most precise measurements of $M_X$ have been made by the 
CDF, Belle, LHCb, and Babar collaborations
\cite{Aaltonen:2009vj,Choi:2011fc,Aaij:2011sn,Aubert:2008gu}.
The PDG average for the mass is $M_X = 3871.68 \pm 0.17$~MeV \cite{Beringer:1900zz}.  
We denote the binding energy relative to the $D^{*0} \bar D^0$ threshold
by $E_X = (M_{*0} + M_0) - M_X$.  Using the PDG averages for
$M_0$ and $M_{*0} - M_0$, we obtain the binding energy
\begin{equation}
E_X =  0.26 \pm 0.39~{\rm MeV}.
\label{EX}
\end{equation}
More precise measurements of $M_0$ by the LHCb collaboration \cite{Aaij:2013uaa}
and by an analysis of data from the CLEOc collaboration \cite{Tomaradze:2012iz}
have further decreased the uncertainty in $E_X$,
reinforcing the conclusion that $X(3872)$ 
is extremely close to the  $D^{*0} \bar D^0$ threshold.
The best experimental upper bound on $\Gamma_X$
comes from measurements in the $J/\psi\, \pi^+ \pi^-$ decay channel
by the Belle collaboration \cite{Choi:2011fc}:
\begin{equation}
\Gamma_X <  1.2~{\rm MeV~~~(90\%~CL)}.
\label{GammaX}
\end{equation}
A theoretical lower bound is provided by the width of the constituent $D^{*0}$:
$\Gamma_X > 0.066$~MeV.

\section{Low-energy Scattering}
\label{sec:DDscat}

In this section, we derive the low-energy scattering amplitudes 
for the three coupled channels consisting of 
neutral and charged charm mesons and $J/\psi\, \omega$.
We then write down simpler scattering amplitudes for the charm mesons only 
in which the effects of the $J/\psi\, \omega$ channel
are taken into account implicitly through one of the scattering parameters.
Finally we write down scattering amplitudes for the charm mesons  
that take into account the $\chi_{c1}(2P)$ resonance.

\subsection{Explicit $\bm{J/\psi\, \omega}$ channel}
\label{sec:DDscatExp}

The low-energy scattering amplitudes $f_{ij}(E)$ 
from S-wave contact interactions between
the three coupled channels defined in Eqs.~(\ref{ch012}) 
can be expressed as a $3 \times 3$ matrix:
\begin{equation}
f(E) =  \left[ - G + K(E) \right]^{-1},
\label{f-GK}
\end{equation}
where $K(E)$ is defined in Eq.~(\ref{Kmatrix})
and $G$ is a symmetric $3 \times 3$ matrix of coupling constants.
Imposing the constraints from isospin symmetry,
this matrix has the form
\begin{eqnarray}
G =  
\Pi_0 \left( 
\begin{array}{c c c}
\gamma_0 &    0     &  \gamma_X  \\
    0   & \gamma_0  & -\gamma_X  \\
\gamma_X & -\gamma_X &  \gamma_V 
\end{array} 
\right) \Pi_0 
+ \gamma_1~\Pi_1,
\label{Gmatrix}
\end{eqnarray}
where $\Pi_0$ and $\Pi_1$ are the isospin projection matrices 
defined in Eqs.~(\ref{ispin01}) and 
$\gamma_0$, $\gamma_1$, $\gamma_V$, and $\gamma_X$ are constants
with dimensions of momentum.
The matrix of amplitudes in Eq.~(\ref{f-GK}) satisfies the 
Lippmann-Schwinger equation:
\begin{equation}
f(E) =
- G^{-1} + G^{-1} \, K(E) \, f(E).
\label{LSeq}
\end{equation}
This can be verified by inserting Eq.~(\ref{f-GK}) for $f(E)$,
multiplying on the left by $G$, and multiplying on the right by $- G + K(E)$, 
in which case it reduces to a trivial identity.
The explicit expressions for the scattering amplitudes $f_{ij}(E)$ 
in Eq.~(\ref{f-GK}) are
\begin{subequations}
\begin{eqnarray}
f_{00} &=&  
[ (-\gamma_0 -\gamma_1 + 2 \kappa_1) (-\gamma_V + \kappa_{\psi \omega}) - 2 \gamma_X^2 ]/D,
\label{f00}
\\ 
f_{01} &=&  
[ (\gamma_1 -\gamma_0) (-\gamma_V + \kappa_{\psi \omega}) - 2 \gamma_X^2 ]/D,
\label{f01}
\\ 
f_{11} &=&  
[ (-\gamma_0 -\gamma_1 + 2 \kappa) (-\gamma_V + \kappa_{\psi \omega}) - 2 \gamma_X^2 ]/D,
\label{f11}
\\ 
f_{02} &=&  
2 (-\gamma_1 + \kappa_1) \gamma_X/D,
\label{f02}
\\ 
f_{12} &=&  
-2 (-\gamma_1 + \kappa) \gamma_X/D,
\label{f12}
\\ 
f_{22} &=&  
[ 2 \gamma_0 \gamma_1 - (\gamma_0 +\gamma_1) (\kappa_1 + \kappa) 
+ 2 \kappa_1 \kappa ]/D ,
\label{f22}
\end{eqnarray}
\label{f-entries}
\end{subequations}
where the denominator is 
\begin{equation}
D = [2 \gamma_0 \gamma_1 - (\gamma_0 +\gamma_1) (\kappa_1 + \kappa) + 2 \kappa_1 \kappa] 
    (-\gamma_V + \kappa_{\psi \omega}) 
- 2 (-2\gamma_1 + \kappa_1 + \kappa) \gamma_X^2.
\label{Den}
\end{equation}
For energies above the appropriate thresholds,
the nonrelativistically normalized T-matrix elements $T_{ij}(E)$ 
for scattering between the three channels are given by the matrix 
\begin{equation}
T(E) =  2 \pi \mu^{-1/2} \, f(E) \, \mu^{-1/2},
\label{T-f}
\end{equation}
where $\mu$ is the diagonal matrix of reduced masses.

The imaginary parts of the scattering amplitudes $f_{ij}(E)$ 
in Eq.~(\ref{f-GK}) can be expressed as
\begin{equation}
{\rm Im}f(E) =  
f(E) \left[ {\rm Im}G - {\rm Im}K(E) \right] f(E)^*,
\label{Imf-GK}
\end{equation}
The T-matrix elements for elastic scattering 
between the three coupled channels are exactly unitary
if the constants $\gamma_0$, $\gamma_1$, $\gamma_V$, and $\gamma_X$ 
in Eq.~(\ref{Gmatrix}) are real and if the widths 
$\Gamma_{*0}$,  $\Gamma_{*1}$, and $\Gamma_\omega$ 
in Eqs.~(\ref{kappa012}) are set to zero.
In this case, the imaginary part of $f_{ij}(E)$ 
is nonzero only if the energy $E$ exceeds one of the thresholds
$0$, $\delta_1$, and $\delta_{\psi \omega}$.
The effects of additional inelastic scattering channels
can be taken into account through the analytic continuation 
of the parameters \cite{Braaten:2007dw,Braaten:2007ft}.
The dominant effects of inelastic scattering channels 
that correspond to decay products of
$D^{*0} \bar D^0$, $D^{*+} D^-$, and $J/\psi\, \omega$,
such as $D^0 \bar D^0 \pi^0$, $D^+ D^- \pi^0$, 
and $J/\psi\, \pi^+ \pi^- \pi^0$
are taken into account through the widths 
$\Gamma_{*0}$,  $\Gamma_{*1}$, and $\Gamma_\omega$ 
in $\kappa$, $\kappa_1$, and $\kappa_{\psi \omega}$.
The dominant effects of other inelastic scattering channels
can be taken into account through the coupling constants 
$\gamma_0$, $\gamma_1$, $\gamma_V$, and $\gamma_X$,
which can have positive imaginary parts. 
For example, the isospin-1 decay mode $J/\psi\, \pi^+ \pi^-$,
in which $\pi^+ \pi^-$ is dominated by the $\rho^0$ resonance,
can be taken into account through the positive imaginary part of 
$\gamma_1$.

As the energy $E$ approaches the $D^{*0} \bar D^0$ threshold at $E=0$,
the elastic scattering amplitude for $D^{*0} \bar D^0$ 
must approach the universal expression \cite{Braaten:2004rn}
\begin{equation}
f_{00}(E) \longrightarrow
 \frac{1}{-\gamma +\kappa(E)}.
\label{f-gamma}
\end{equation}
It is easy to identify $\gamma$ by exploiting the fact that the 
denominator $D$ in Eq.~(\ref{Den}) is linear in $\kappa$.  
Since $|E| \ll \delta_1,\delta_{\psi \omega}$,
we can set $E=0$ in $\kappa_1$ and $\kappa_{\psi \omega}$.
The resulting expression for the inverse scattering length is
\begin{equation}
\gamma =
\frac{ 2 \gamma_0 \gamma_1 - (\gamma_0 +\gamma_1) \kappa_1(0) 
+ 2 [2\gamma_1 - \kappa_1(0)] \gamma_X^2/[-\gamma_V + \kappa_{\psi \omega}(0)]}
{\gamma_0 + \gamma_1 - 2 \kappa_1(0) 
+ 2 \gamma_X^2/[-\gamma_V + \kappa_{\psi \omega}(0)]}.
\label{gamma}
\end{equation}
The binding energy and the width of the $X(3872)$ 
are determined by the real and imaginary parts of $\gamma$.
This puts two constraints on the real parts 
and the small imaginary parts of the 
four parameters $ \gamma_0$, $ \gamma_1$, $\gamma_V$, and $\gamma_X$.

\subsection{Implicit $\bm{J/\psi\, \omega}$ channel}
\label{sec:DDscatImp}

If $\gamma_X = 0$ or if $|\gamma_V|$ is much larger than 
$|\kappa_{\psi \omega}(E)|$, 
the $J/\psi\, \omega$ channel decouples from 
the charm meson channels. 
The $J/\psi\, \omega$ scattering amplitude $f_{22}$ 
in Eq.~(\ref{f22}) reduces to 
$1/(-\gamma_V + \kappa_{\psi \omega})$.
The scattering amplitudes 
for the 0 and 1 channels reduce to the scattering amplitudes 
for charm mesons derived in Ref.~\cite{Braaten:2007ft}:
\begin{subequations}
\begin{eqnarray}
f_{00} &=&  
(-\gamma_0 -\gamma_1 + 2 \kappa_1)/D',
\label{fBL00}
\\ 
f_{01} &=&  
(\gamma_1 -\gamma_0)/D',
\label{fBL01}
\\ 
f_{11} &=&  
(-\gamma_0 -\gamma_1 + 2 \kappa)/D',
\label{fBL11}
\end{eqnarray}
\label{fBL-entries}
\end{subequations}
where the denominator is 
\begin{equation}
D' = 2 \gamma_0 \gamma_1 - (\gamma_0 +\gamma_1) 
(\kappa_1 + \kappa) + 2 \kappa_1 \kappa.
\label{DenBL}
\end{equation}
These are the appropriate scattering amplitudes if 
the $X(3872)$ resonance is generated dynamically 
by attractive interactions between the charm mesons.

The charm meson scattering amplitudes $f_{00}$, $f_{01}$,
and $f_{11}$ in Eqs.~(\ref{f-entries}),
which take into account the $J/\psi\, \omega$ channel explicitly,
can be obtained exactly from the amplitudes in Eqs.~(\ref{fBL-entries})
by making the substitution
\begin{equation}
\gamma_0 \longrightarrow  
\gamma_0 + \frac{2 \gamma_X^2}{-\gamma_V + \kappa_{\psi \omega}(E)}.
\label{gamma0-psiomega}
\end{equation}
The second term on the right side is the product of the 
$J/\psi\, \omega$ scattering amplitude and transition amplitudes
proportional to $\gamma_X$.
Thus the only effect of the $J/\psi\, \omega$ channel on scattering 
between the charm mesons is to resolve the isospin-0 inverse
scattering length into an energy-dependent term from transitions to 
$J/\psi\, \omega$ and a constant $\gamma_0$ that takes into account
shorter-distance effects.

\subsection{$\bm{\chi_{c1}(2P)}$ resonance}
\label{sec:DDscatY}

The $\chi_{c1}(2P)$ charmonium state has quantum numbers $1^{++}$.
If its mass is close enough to that of the $X(3872)$ resonance,
it can have a significant effect on the charm meson scattering amplitudes 
near the $D^{*0} \bar D^0$ threshold.
It could be responsible for generating the $X(3872)$ resonance
or it could be a separate resonance with quantum numbers $1^{++}$.
If the $\chi_{c1}(2P)$ is a separate resonance from the $X(3872)$,
it is expected to be higher in mass.
One former candidate for the $\chi_{c1}(2P)$ is a state labelled
$X(3915)$ by the Particle Data Group \cite{Beringer:1900zz}.
It was discovered by the Babar collaboration
through $B$ decays into $X(3915)+K$ in the decay channel
$X(3915) \to J/\psi\, \omega$ \cite{Aubert:2007vj},
which implies that its charge conjugation is $C=+$.
Its properties were measured more accurately in Ref.~\cite{delAmoSanchez:2010jr}. 
The $X(3915)$ was also observed by the Belle collaboration
in the production channel $\gamma \gamma \to X(3915)$ \cite{Uehara:2009tx},
which would have excluded $1^{++}$,
but that observation could also be attributed instead to 
the nearby charmonium state $\chi_{c2}(2P)$ at 3927~MeV.
However a recent analysis by the Babar collaboration of 
$\gamma \gamma \to X(3915) \to J/\psi\, \omega$
determined the spin-parity to be $J^P = 0^+$ \cite{Lees:2012xs}.
This excludes $X(3915)$ as a candidate for $\chi_{c1}(2P)$.
At this point, there is no well-established resonance besides 
the $X(3872)$ that might be identified with $\chi_{c1}(2P)$.
We will however for completeness consider the possibility of 
a separate $1^{++}$ resonance with mass above 3872~MeV.

The coupled-channel problem for low-energy S-wave interactions 
of neutral and charged charm meson pairs 
with a $1^{++}$ charmonium resonance was solved 
in Ref.~\cite{Artoisenet:2010va}.
The scattering amplitudes are those for charm mesons in
Eqs.~(\ref{fBL-entries}) with the substitution
\begin{equation}
\gamma_0 \longrightarrow  
\left( \frac{1}{\gamma_0} + \frac{g^2}{E - \nu} \right)^{-1}
= \frac{\gamma_0 (E - \nu)}{E - \nu + g^2 \gamma_0}.
\label{gamma0-Y}
\end{equation}
These scattering amplitudes were also studied in Ref.~\cite{Hanhart:2011jz}.
They are exactly unitary for real values of the four parameters
$\gamma_0$, $\gamma_1$, $\nu$, and $g$. 
The combination $\nu - g^2 \gamma_0$ can be identified as the energy of the 
$\chi_{c1}(2P)$ resonance.
In Ref.~\cite{Artoisenet:2010va}, charmonium phenomenology 
was used to obtain the estimate $g = 0.4$ for the coupling constant.
The methods of Ref.~\cite{Artoisenet:2010va} could be extended 
to the case with a third scattering channel $J/\psi\, \omega$
that also couples to the $\chi_{c1}(2P)$.

\section{Line shapes of $\bm{X(3872)}$}
\label{sec:lineshape}

In this section, we present line shapes for the $X(3872)$ resonance 
in the $J/\psi\, \pi^+ \pi^- \pi^0$ channel.
We first discuss the short-distance factors in a
factorization formula for the line shapes.
We give an expression for the line shape
in which the $J/\psi\, \omega$ channel is taken into account explicitly.
We then give an expression for the line shape
in which the effects of the $J/\psi\, \omega$ channel
are taken into account implicitly through the scattering parameter $\gamma_0$.
Finally we give an expression for the line shape
in which the $\chi_{c1}(2P)$ resonance is taken into account.

\subsection{Short-distance factors}
\label{sec:lineshapeSD}

For a production process that involves an energy transfer that is large 
compared to the low-energy scales $\delta_1$ and $\delta_{\psi \omega}$
set by the differences between the thresholds,
the inclusive production rate summed over all resonant final states $X$
satisfies a factorization formula \cite{Braaten:2005jj}.
If the production process is a decay, such as $B \to K+X$ or $B \to K^*+X$,
the differential decay rate can be expressed as
\begin{equation}
d\Gamma = \mbox{$\sum$}_{ij} \Gamma_{ij} \,  {\rm Im} f_{ij}(E) \, dE,
\label{dGamma/dE}
\end{equation}
where the  $\Gamma_{ij}$ are short-distance factors 
that are insensitive to the resonance energy $E$.

The expression for the matrix Im$f(E)$ in Eq.~(\ref{Imf-GK})
can be used to decompose the differential decay rate in Eq.~(\ref{dGamma/dE})
into contributions proportional to the imaginary parts 
of the scattering parameters, which appear in the matrix $G$,
and the imaginary parts 
of the threshold factors, which appear in the matrix $K$.
The short-distance factors $\Gamma_{ij}$ in Eq.~(\ref{dGamma/dE})
are entries of a positive-definite hermitian matrix.
They can be expressed as sums with positive weights 
of terms of the form $C_{k,i} (C_{k,j})^*$,
where $C_{k,i}$ is a short-distance amplitude for
the creation of a pair of particles in the channel $i$.
The sum is over transition channels $k$ from the initial state
to the additional final-state particles besides those in the resonance channel.
Constraints on these short-distance amplitudes from the symmetries of QCD 
imply constraints on the short-distance factors $\Gamma_{ij}$.
The decay $B^+ \to K^+ + X$ is particularly simple,  because
the transition $B^+ \to K^+$ between the two spin-0 particles
has a single short-distance amplitude $C_i$.
The short-distance factors can therefore be expressed as
\begin{equation}
\Gamma_{ij} = C_i (C_j)^*.
\label{Gamma-C}
\end{equation}
The isospin symmetry of decays that proceed 
at the quark level through the heavy quark decay $b \to c \bar c s$
relates the short-distance amplitudes for $B^0 \to K^0 + X$
and $B^+ \to K^+ + X$:
\begin{eqnarray}
\left( 
\begin{array}{c}
C_{0}   \\ C_{1}   \\ C_{2}   
\end{array} \right)_{B^0 \to K^0} =
\left( 
\begin{array}{c}
C_{1}   \\ C_{0} \\ C_{2}
\end{array} \right)_{B^+ \to K^+} .
\label{isospinsym}
\end{eqnarray}
The constraints  on the coefficients $C_0$ and $C_1$  were derived previously 
\cite{Braaten:2007ft,Artoisenet:2010va}.
The constraint on the coefficients $C_2$ follows from the 
equality of the short-distance amplitudes for  
$B^0 \to K^0 + (J/\psi\, \omega)$ and $B^+ \to K^+ + (J/\psi\, \omega)$, 
which is required by isospin symmetry.
Using the simple form for the short-distance factors for $B \to X$
in Eq.~(\ref{Gamma-C}), the differential decay rate 
in Eq.~(\ref{dGamma/dE}) reduces to
\begin{equation}
d\Gamma = \mbox{$\sum$}_{ijkl} [C_i f_{ik}(E)]\, [C_j f_{jl}(E)]^*\,  
\left[ {\rm Im} G_{kl} - {\rm Im} K_{kl}(E) \right] dE.
\label{dGamma/dE:BtoK}
\end{equation}

We can obtain order-of-magnitude estimates for
the ratios of $|C_0|^2$, $|C_1|^2$, and $|C_2|^2$ from
measured partial widths of $B$ into $K$ plus appropriate pairs of mesons.
The decay amplitude into three mesons, such as  $D^{*0} \bar D^0 K^+$,
is a function of two Lorentz invariants whose range extends over the Dalitz plot
for the three mesons.
The amplitudes $C_0$, $C_1$, and $C_2$ are the short-distance factors 
of the decay amplitudes in the corner of the Dalitz plot corresponding 
to the threshold for the two mesons other than $K$.  
Our estimates of their ratios are based on the assumption 
that the short-distance factors 
do not vary dramatically over the Dalitz plot.  The partial
widths of $B$ into $K$ plus pairs of charm mesons have been measured
by the Babar collaboration in Ref.~\cite{delAmoSanchez:2010pg}.  The
partial widths into $K$ plus $J/\psi\, \omega$ were measured by the
Babar collaboration in Ref.~\cite{delAmoSanchez:2010jr}.  Using the
data from $B^+$ decays, we estimate $|C_1|^2/|C_0|^2$ by dividing the
sum of the partial widths for $D^{*+} D^-$ and $D^+\, D^{*-}$ by the
sum of the partial widths for $D^{*0} \bar D^0$ and $D^0 \bar D^{*0}$.
Using the data from $B^0$ decays, we must interchange the numerator
and denominator.  The resulting estimates for $|C_1|^2/|C_0|^2$ are
0.14 from $B^+$ decays and 0.17 from $B^0$ decays.  Using the data
from $B^+$ decays, we estimate $|C_2|^2/|C_0|^2$ by dividing the
partial width for $J/\psi\, \omega$ by the sum of the partial widths
for $D^{*0} \bar D^0$ and $D^0 \bar D^{*0}$.  Using the data from
$B^0$ decays, we must replace the denominator by the sum of the
partial widths for $D^{*+} D^-$ and $D^+\, D^{*-}$.  The resulting
estimates for $|C_2|^2/|C_0|^2$ are 0.037 from $B^+$ decays and 0.036
from $B^0$ decays.  These estimates suggest that the short-distance
production rates for $J/\psi\, \omega$ and for $D^{*+} D^-$ and $D^+\,
D^{*-}$ are smaller than those for $D^{*0} \bar D^0$ and $D^0 \bar
D^{*0}$ by factors of about 30 and 6.5, respectively.

The suppression of $C_2$ relative to $C_0$ and $C_1$ does not necessarily imply 
that the $C_2$ term in the resonance factor in Eq.~(\ref{dGamma-psiomega})
can be neglected.  The $C_0$ and $C_1$ terms in the resonance factor 
are multiplied by $\gamma_X$, 
which is an amplitude for a transition between $J/\psi\, \omega$ 
and a pair of charm mesons.  Since this process involves a rearrangement of 
constituent charm quarks between the two mesons, $\gamma_X$ could 
provide a sufficient suppression factor to make the $C_0$ and $C_1$ terms
comparable in strength to the $C_2$ term.
For production of the resonance in other channels, 
such as $J/\psi\, \rho$ and $D^0 \bar D^0 \pi^0$,
the $C_2$ term in the resonance factor should be completely negligible.

\subsection{Explicit $\bm{J/\psi\, \omega}$ channel}
\label{sec:lineshapeExp}

\begin{figure}[t]
\vspace*{-0.0cm}
\includegraphics*[width=6cm,angle=0,clip=true]{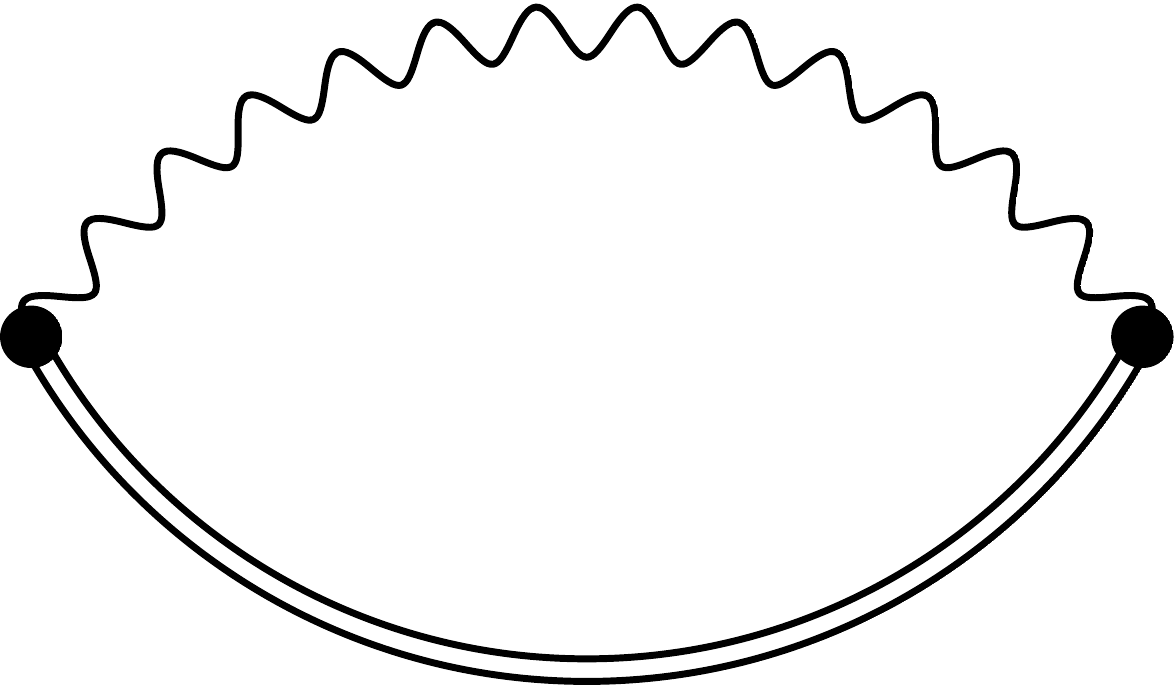}
\hspace{1cm}
\includegraphics*[width=6cm,angle=0,clip=true]{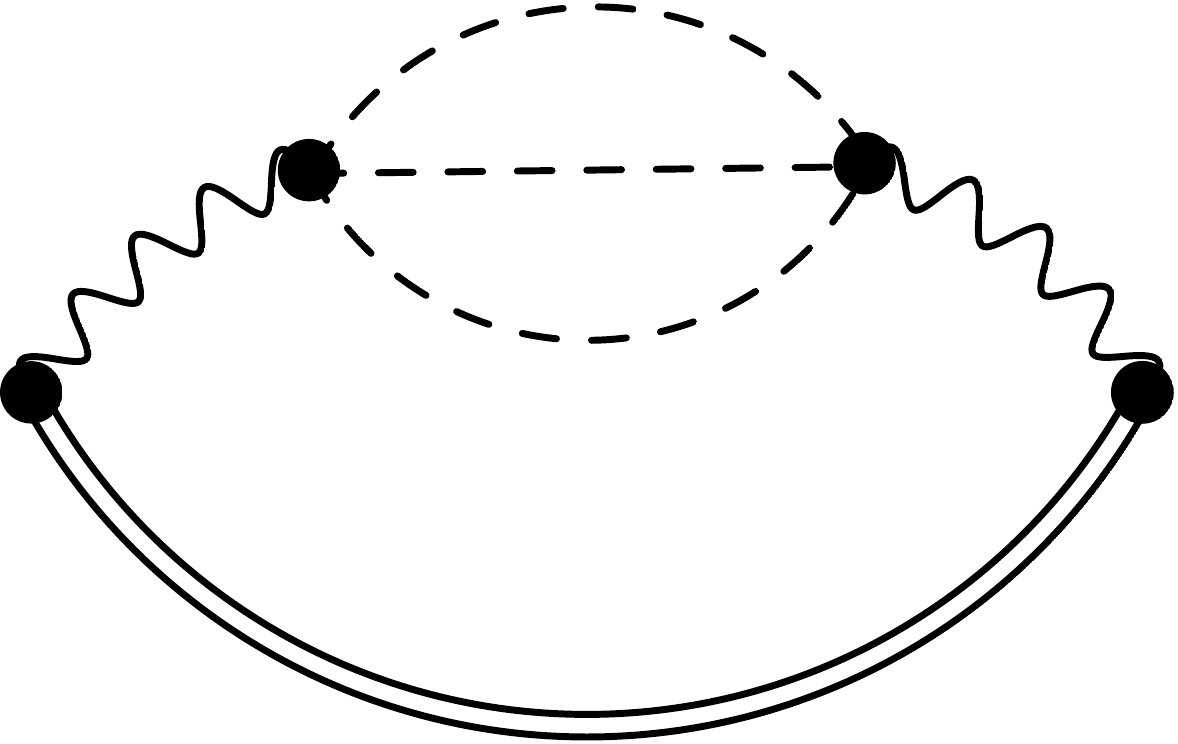}
\vspace*{-0.0cm}
\caption{(Color online) 
Diagrams for the propagation of $J/\psi\, \omega$ between contact interactions:
(a) simple bubble diagram with a $J/\psi\, \omega$ cut,
(b) diagram with a $J/\psi\, \pi^+\pi^-\pi^0$ cut.
The $J/\psi$, $\omega$, and pions are represented by
double solid, wavy, and dashed lines, respectively.
}
\label{fig:psiomega}
\end{figure}

The inclusive differential decay rate for $B \to X + K$ in Eq.~(\ref{dGamma/dE:BtoK})
can be partially resolved into contributions from individual resonant states by
inserting the expressions for ${\rm Im} f_{ij}(E)$ in Eq.~(\ref{Imf-GK}).
The imaginary part of $K_{22} = \kappa_{\psi \omega}$ comes from
cutting the bubble diagram in Fig.~\ref{fig:psiomega}(a),
in which $J/\psi$ and $\omega$ propagate 
between points where they are created and annihilated.
The term in the differential rate proportional to Im$\kappa_{\psi \omega}$
therefore represents the contribution from the final state $J/\psi\, \omega$
or from decay products of this pair of particles:
\begin{equation}
d\Gamma[J/\psi\, \omega] = 
\big| \mbox{$\sum$}_{i=0,1,2} C_i f_{i2}(E) \big|^2 
\left( - {\rm Im} \kappa_{\psi \omega}(E) \right) \, dE,
\label{dGamma-psiomega}
\end{equation}
where $f_{i2}(E)$ are the scattering amplitudes in 
Eqs.~(\ref{f-entries}).
The imaginary part of the function $\kappa_{\psi \omega}(E)$  
in Eq.~(\ref{kappaV}) can be  expressed in analytic form:
\begin{equation}
- {\rm Im} \kappa_{\psi \omega}(E)  =
\mu_{\psi \omega} ^{1/2} 
\left( \sqrt{(E - \delta_{\psi \omega})^2 + \Gamma_\omega^2/4} 
+ E - \delta_{\psi \omega} \right)^{1/2}.
\label{ImkappaV}
\end{equation}
If $E$ is above the threshold $\delta_{\psi \omega}$ 
by much more than $ \Gamma_\omega/2$,
the expression in Eq.~(\ref{ImkappaV}) reduces to
$[2 \mu_{\psi \omega} (E - \delta_{\psi \omega})]^{1/2}$.
In this region of $E$, Eq.~(\ref{dGamma-psiomega}) 
is the differential rate for producing $J/\psi$ and $\omega$ 
on their mass shells.
If $E$ is below the threshold $\delta_{\psi \omega}$ 
by much more than $\Gamma_\omega/2$,
the expression in Eq.~(\ref{ImkappaV}) reduces to
$( \frac18 \mu_{\psi \omega} / 
|E - \delta_{\psi \omega}| )^{1/2} \Gamma_\omega$.
In this region of $E$, Eq.~(\ref{dGamma-psiomega}) 
is the differential rate for producing $J/\psi$ 
plus the decay products of a virtual $\omega$, such as $\pi^+ \pi^- \pi^0$.
The contribution from the specific decay channel 
$J/\psi\, \pi^+ \pi^- \pi^0$ can be obtained from  
Eq.~(\ref{dGamma-psiomega}) by multiplying by the branching fraction
$B_{\omega\to \pi\pi\pi} \approx 89\%$.

As the energy $E$ approaches the $D^{*0} \bar D^0$ threshold,
the elastic scattering amplitude for $D^{*0} \bar D^0$ 
approaches the universal expression
in Eq.~(\ref{f-gamma}).  The pole at $\kappa(E) = \gamma$ 
arises from the denominator 
$D(E)$, which is a common factor in all the scattering amplitudes 
in  Eq.~(\ref{f-entries}).
Thus all the scattering amplitudes have that same energy dependence 
near the $D^{*0} \bar D^0$ threshold.  
The linear combination of scattering amplitudes 
that appears in the resonance factor in Eq.~(\ref{dGamma-psiomega})
has the behavior
\begin{equation}
\mbox{$\sum$}_{i} C_i f_{i2}(E) \longrightarrow
\frac{ 2 [\gamma_1 - \kappa_1]  \gamma_X C_0 - 2 \gamma_1  \gamma_X C_1
- [2 \gamma_1 \gamma_0 - (\gamma_1 + \gamma_0) \kappa_1] C_2}
{[\gamma_0 + \gamma_1 - 2 \kappa_1] [-\gamma_V + \kappa_{\psi \omega}] 
+ 2 \gamma_X^2}\, 
 \frac{1}{-\gamma +\kappa(E)},
\label{sumCf-gamma}
\end{equation}
where $\kappa_1$ and $\kappa_{\psi \omega}$ are evaluated at $E=0$.
Thus the resonance factor in Eq.~(\ref{dGamma-psiomega}) 
has the simple universal form
$|-\gamma +\kappa(E)|^{-2}$ at energies $E$ small compared to the 
thresholds $\delta_1$ and $\delta_{\psi \omega}$,
which are both approximately 8~MeV.

\subsection{Implicit $\bm{J/\psi\, \omega}$ channel}
\label{sec:lineshapeImp}

If the short-distance factor $C_2$ for the production of 
$J/\psi\, \omega$ is sufficiently small, it is not essential to take the 
$J/\psi\, \omega$ channel into account explicitly.  
It can be taken into account implicitly through the 
isospin-0 inverse scattering length $\gamma_0$.
In the factorization formula in Eq.~(\ref{dGamma/dE:BtoK}),
$\gamma_0$ appears in the coupling constant matrix $G$.
The contribution to Im$(\gamma_0)$ from the 
$J/\psi\, \omega$ channel
can be deduced from the substitution for $\gamma_0$ 
given in Eq.~(\ref{gamma0-psiomega}):
\begin{equation}
\left( {\rm Im} \gamma_0 \right)_{J/\psi\, \omega} \longrightarrow
\frac{2 \gamma_X^2}{|-\gamma_V+\kappa_{\psi \omega}(E)|^2}
 \left( - {\rm Im} \kappa_{\psi \omega}(E) \right),
\label{Imgamma0-psiomega}
\end{equation}
where the terms with ${\rm Im} \gamma_V$ and ${\rm Im} \gamma_X$ 
have been dropped 
because they do not contribute to the $J/\psi\, \omega$ final state.
Our final result for the decay rate into the $J/\psi\, \omega$ channel is
\begin{equation}
d\Gamma[J/\psi\, \omega] = 
\big| \mbox{$\sum$}_{i=0,1} C_i (f_{i0}(E) - f_{i1}(E)) \big|^2
\frac{\gamma_X^2}{|-\gamma_V+\kappa_{\psi \omega}(E)|^2} 
\left( - {\rm Im} \kappa_{\psi \omega}(E) \right) \, dE,
\label{dGamma-psiomegaBL}
\end{equation}
where $f_{i0}(E)$ and $f_{i1}(E)$ are the scattering amplitudes in 
Eqs.~(\ref{fBL-entries}) with the substitution for $\gamma_0$  in
Eq.~(\ref{gamma0-psiomega}).
This result can also be obtained from the expression in
Eq.~(\ref{dGamma-psiomega}) in which the $J/\psi\, \omega$ channel 
is taken into account explicitly by setting $C_2=0$.

\subsection{$\bm{\chi_{c1}(2P)}$ resonance}
\label{sec:lineshapechi}

The scattering amplitudes for charm mesons that take into account 
the possibility that the $1^{++}$ charmonium resonance 
$\chi_{c1}(2P)$ is near the $X(3872)$ 
are given by Eqs.~(\ref{fBL-entries})
with the substitution for $\gamma_0$ in Eq.~(\ref{gamma0-Y}).
The corresponding expressions for the line shapes
were derived in Ref.~\cite{Artoisenet:2010va}.
They take into account the short-distance production of $\chi_{c1}(2P)$
as well as charm mesons.
The contributions from the imaginary parts of 
$\gamma_0$, $\gamma_1$, and $\nu$ were taken into account, 
but the coupling constant $g$ was assumed to be real.
The line shapes were used to carry out a phenomenological analysis 
of the $J/\psi\, \pi^+\pi^-$, $D^0 \bar D^0 \pi^0$,
and $D^0 \bar D^0 \gamma$ channels.

We can use the results in Ref.~\cite{Artoisenet:2010va} to write down 
an expression for the line shape in the 
$J/\psi\, \omega$ channel.
For simplicity, we ignore the possibility of the 
short-distance production of $\chi_{c1}(2P)$.
The inclusive line shape in Eq.~(\ref{dGamma/dE:BtoK})
for isospin-0 channels produced by $B \to X+K$ reduces to
\begin{equation}
d\Gamma[{\rm isospin~0}] = 
\big| \mbox{$\sum$}_{i=0,1} C_i (f_{i0}(E) - f_{i1}(E)) \big|^2~ 
{\rm Im} 
\left( 1/\gamma_0 + g^2/(E-\nu) \right)^{-1} \, dE,
\label{dGamma-psiomegachi}
\end{equation}
where $f_{i0}(E)$ and $f_{i1}(E)$ are the scattering amplitudes in 
Eqs.~(\ref{fBL-entries}) with the substitution for $\gamma_0$ in
Eq.~(\ref{gamma0-Y}).
The expression for the imaginary part in Eq.~(\ref{dGamma-psiomegachi})
that corresponds to cutting rules is
\begin{eqnarray}
{\rm Im} 
\left( \frac{1}{\gamma_0} + \frac{g^2}{E-\nu} \right)^{-1} =
\frac{1}{|E-\nu+g^2 \gamma_0|^2}
\Bigg( |E-\nu|^2 {\rm Im}(\gamma_0)
+ |\gamma_0|^2 |g|^2 {\rm Im}(-\nu)
\nonumber \\
- 2|\gamma_0|^2 {\rm Re}[g(E-\nu^*)] {\rm Im}(g) \Bigg).
\label{Imgammachi}
\end{eqnarray}
The first two terms in the parentheses
can be interpreted as contributions from
inelastic charm meson scattering and from $\chi_{c1}(2P)$ decay,
respectively.  The imaginary parts of $\gamma_0$ and $-\nu$ must be
positive.  The third term in Eq.~(\ref{Imgammachi}) can be attributed
to interference between inelastic charm meson scattering and
$\chi_{c1}(2P)$ decay.  
Positivity of the line shape for all energies $E$ requires 
$({\rm Im} g)^2 \le {\rm Im}(-\nu){\rm Im}(\gamma_0)/|\gamma_0|^2$.  
The line shape in the $J/\psi\, \pi^+ \pi^-\pi^0$ channel 
can be obtained by inserting
Eq.~(\ref{Imgammachi}) into Eq.~(\ref{dGamma-psiomegachi}), by
replacing the imaginary parts of $\gamma_0$, $-\nu$, and $g$ by the
$J/\psi\, \omega$ channel contributions to the imaginary parts, and by
multiplying by the branching fraction $B_{\omega\to \pi\pi\pi}$.

\section{Three-pion invariant mass distribution}
\label{sec:3pionmass}

In this section, we derive a simple expression for the 
distribution of the invariant mass $M_{3\pi}$ of the three pions
in the decay channel $J/\psi\, \pi^+ \pi^- \pi^0$.
We describe the results of the Babar analysis of the $M_{3\pi}$
distribution and point out that the probability for the 
quantum numbers $1^{++}$ for the $X(3872)$ was underestimated.
We describe previous theoretical analyses of the $M_{3\pi}$
distribution, which ignored the effects of experimental resolution.
We also study the effects of scattering on the $M_{3\pi}$
distribution.

\subsection{$\bm{M_{3 \pi}}$ distribution}

The differential decay rate in Eq.~(\ref{dGamma-psiomega}) 
is differential only in the energy $E$.  We proceed to derive 
the $M_{3 \pi}$ distribution within the same framework.
The factor of $- {\rm Im} \kappa_{\psi \omega}(E)$  in Eq.~(\ref{dGamma-psiomega})
comes from cutting the bubble diagram in Fig.~\ref{fig:psiomega}(a).
In the case of the final state $J/\psi\, \pi^+ \pi^- \pi^0$,
we can obtain an expression that is differential in additional variables
by replacing that cut diagram by a cut of the diagram in Fig.~\ref{fig:psiomega}(b)
in which the cut passes through $J/\psi\, \pi^+ \pi^- \pi^0$.  
The distribution of the invariant mass $M_{3\pi}$ 
of the pions would be obtained by integrating over all the other pion variables 
besides $M_{3\pi}$.  This distribution can also be obtained more simply from 
the $J/\psi\, \omega$ cut diagram in Fig.~\ref{fig:psiomega}(a).  
In the nonrelativistic limit,
the relation between $M_{3\pi}$, the total energy $M= (M_{*0} + M_0)+E$,  
and the relative momentum $q$ of the $J/\psi$ or the virtual $\omega$ is
\begin{equation}
M = M_{3\pi}  + M_\psi + q^2/(2 \mu_{\psi \omega}).
\label{M-M3pi}
\end{equation}
If the width of the $\omega$ is included in its propagator,
the momentum integral for the $J/\psi\, \omega$ cut diagram multiplied by 
$4\pi(M_\psi+M_\omega)$ is
\begin{eqnarray}
&& \frac{\pi}{\mu_{\psi \omega}} \int \frac{d^3q}{(2 \pi)^3}  
\frac{\Gamma_\omega}
{| E - \delta_{\psi \omega} -  q^2/(2 \mu_{\psi \omega}) + i \Gamma_\omega/2 |^2}
\nonumber \\
&& \hspace{1cm} 
= \frac{\Gamma_\omega} {2 \pi}
\int_{-\infty}^{M - M_\psi } dM_{3\pi}\, 
\frac{ \sqrt{2 \mu_{\psi \omega} (M - M_\psi - M_{3\pi})}}
{(M_{3\pi} - M_\omega)^2 + \Gamma_\omega^2/4} .
\label{int3pi}
\end{eqnarray}
The lower limit on the integral over $M_{3\pi}$ extends below
the physical lower limit of $3 m_\pi$, but the unphysical region is strongly 
suppressed by the Breit-Wigner factor.
Upon integrating over $M_{3\pi}$, Eq.~(\ref{int3pi})
reproduces the expression for $- {\rm Im} \kappa_{\psi \omega}(E)$
in Eq.~(\ref{ImkappaV}).  An expression for the differential decay rate 
that is differential in both $E$ and $M_{3\pi}$ can therefore be obtained by 
replacing $- {\rm Im} \kappa_{\psi \omega}(E)$ by the integrand 
on the right side of Eq.~(\ref{int3pi}):
\begin{eqnarray}
d\Gamma[J/\psi\, \omega] = 
\big| \mbox{$\sum$}_{i} C_i f_{i2}(E) \big|^2 
\frac{\Gamma_\omega \; q}
{2 \pi [(M_{3\pi} - M_\omega)^2 + \Gamma_\omega^2/4]}\,
dM_{3\pi}\, dM,
\label{dGamma/dEdM}
\end{eqnarray}
where $E = M - (M_{*0} + M_0)$
and $q = [2 \mu_{\psi \omega} (M - M_\psi - M_{3\pi})]^{1/2}$
is the relative momentum of the $J/\psi$ or the virtual $\omega$.
This expression for the differential decay rate is Lorentz invariant.
The dependence on $M_{3\pi}$ is simply the product of $q$
and a Breit-Wigner resonance function.
The differential decay rate into $J/\psi\, \pi^+ \pi^- \pi^0$
can be obtained by multiplying the right side of Eq.~(\ref{dGamma/dEdM})
by the branching fraction $B_{\omega\to \pi\pi\pi}$.

\subsection{Babar data}

\begin{figure}[t]
\vspace*{-0.0cm}
\centerline{\includegraphics*[width=\columnwidth,angle=0,clip=true]{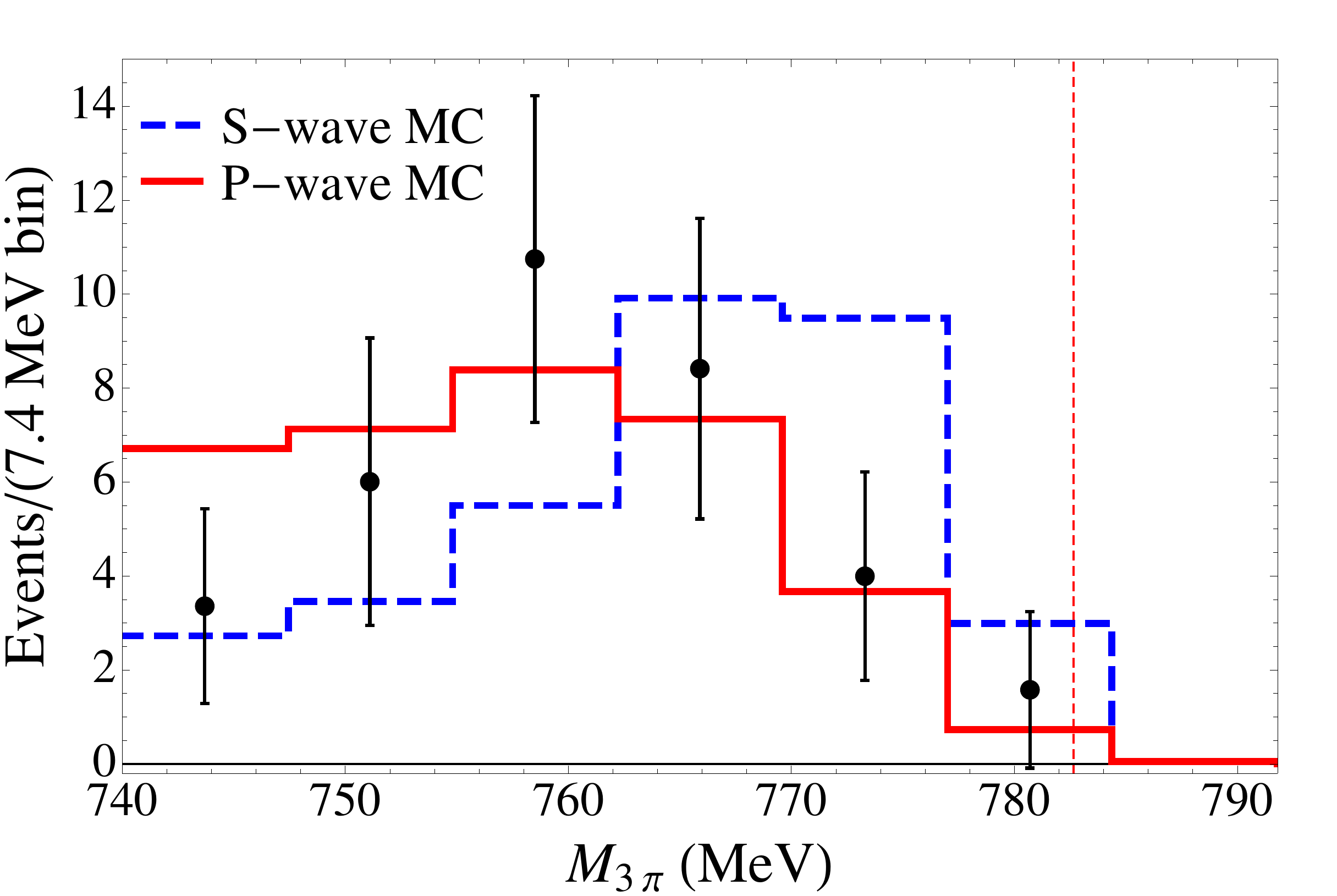}}
\vspace*{-0.0cm}
\caption{(Color online) 
Uncorrected distribution of $M_{3\pi}$ 
integrated over the $J/\psi\, \pi^+\pi^-\pi^0$ 
invariant mass $M$ from 3862.5~MeV to 3882.5~MeV.
The data points are the Babar data from Ref.~\cite{delAmoSanchez:2010jr}.
The histograms are for Monte Carlo events generated by Babar
under the assumption that the coupling of $X(3872)$ to $J/\psi\, \omega$
is S-wave (dashed lines) or P-wave (solid lines) \cite{delAmoSanchez:2010jr}.
The vertical dashed line marks the position 
of the center of the $\omega$ resonance.}
\label{fig:Babar}
\end{figure}

The Babar data on the decay of $X(3872)$ into $J/\psi\, \omega$ that
favors the quantum numbers $2^{-+}$ over $1^{++}$ is the $M_{3 \pi}$
distribution shown in Fig.~\ref{fig:Babar} \cite{delAmoSanchez:2010jr}.
The $J/\psi\, \pi^+\pi^-\pi^0$ invariant mass $M$ is integrated over
the range from 3862.5~MeV to 3882.5~MeV, which extends about 10~MeV
above and below the $D^{*0} \bar D^0$ threshold.  The range of $M_{3
  \pi}$ in Fig.~\ref{fig:Babar} is from 740~MeV to 791.8~MeV, which is
approximately $M_\omega - 5 \Gamma_\omega$ to $M_\omega +
\Gamma_\omega$.  The Babar data in Fig.~\ref{fig:Babar} consists of
$34.0 \pm 6.6$ events including a background of $8.9\pm 1.0$ events.
Also shown in Fig.~\ref{fig:Babar} are histograms of Monte Carlo events
generated by the Babar collaboration under the assumptions that the
coupling of $X$ to $J/\psi\, \omega$ is either S-wave or P-wave.  The
histograms are normalized to 34 events.  Since the P-wave Monte Carlo
gives a better fit to the $M_{3 \pi}$ distribution, the Babar
collaboration concluded that the quantum numbers $2^{-+}$ are favored
over $1^{++}$.

A quantitative measure of the quality of the fit is $\chi^2$ of the
histogram with respect to the 6 nonzero data points.  In
Ref.~\cite{delAmoSanchez:2010jr}, the values of $\chi^2$ per
degree of freedom were given as
\begin{subequations}
\begin{eqnarray}
\label{chi2Babar}
&\chi^2_\text{Babar}/\text{NDF}=
10.17/5  &\hspace{1cm} \text{for S-wave  Monte Carlo}\,,
\\
&\chi^2_\text{Babar}/\text{NDF}=
3.53/5   &\hspace{1cm} \text{for P-wave Monte Carlo}\,.
\end{eqnarray}
\end{subequations}
The probabilities for $\chi^2$ to be larger than these values are 
7.1\% and 61.9\%, respectively.
This seems to indicate that P-wave coupling of $X$ to $J/\psi\, \omega$
(and therefore quantum numbers $2^{-+}$)
is strongly favored over S-wave coupling (and quantum numbers $1^{++}$).
However $\chi^2(N)$ is a function of the normalization $N$ of the histograms.
The values of $\chi^2$ given in Ref.~\cite{delAmoSanchez:2010jr} 
were for histograms normalized to $\bar N = 34$ events,
which is the central value of the sum of the data points.  
Normalizing them in this way is fine for illustrating differences
in their qualitative behavior, as in Fig.~\ref{fig:Babar}.
However it is not appropriate for calculating the $\chi^2$,
because it does not allow for independent fluctuations in the 6 bins.
Instead it requires that any downward fluctuations in some bins
be compensated by upward fluctuations in other bins.
Furthermore there is no guarantee that the probability distribution for $\chi^2(\bar N)$ 
is the standard $\chi^2$ probability distribution.
The quantity that has the probability distribution
of $\chi^2$ for 5 degrees of freedom in the limit of ideal measurements
is $\chi^2(N)$ minimized with respect to $N$.
The minimum $\chi^2$ per degree of freedom for the Babar data is
\begin{subequations}
\begin{eqnarray}
\label{chi2min}
&\chi^2(N_\text{min})/\text{NDF}=
7.49/5  &\hspace{1cm} \text{for S-wave Monte Carlo}\,,
\\
&\chi^2(N_\text{min})/\text{NDF}=
3.25/5   &\hspace{1cm} \text{for P-wave Monte Carlo}\,.
\end{eqnarray}
\end{subequations}
For the S-wave Monte Carlo, the minimum is at
$N_\text{min} = 24.9$ and the probability 
for $\chi^2$ to be larger than the observed value is 18.7\%.
For the P-wave Monte Carlo, $N_\text{min} = 30.3$
and the probability for $\chi^2$ to be larger than the observed value 
is 66.2\%.   While a P-wave coupling is still favored over S-wave, 
it is not favored as strongly as reported
in Ref.~\cite{delAmoSanchez:2010jr}.

In order to compare a theoretical distribution to one that is measured,
it is essential to take into account the experimental resolution.
In the Babar experiment,
the experimental resolution on $M$ is $\sigma_X=6.7$~MeV \cite{delAmoSanchez:2010jr}. 
The experimental resolution on 
$M_{3 \pi}$ was not given in Ref.~\cite{delAmoSanchez:2010jr}.  
A reasonable estimate is the resolution of the $\omega$ mass in a study of the 
decay $\bar B^0 \to D^* \omega \pi^-$, which was 
$\sigma_\omega = 5.6$~MeV \cite{Aubert:2006zb}.
These resolutions can be taken into account by convolving
the distribution in Eq.~(\ref{dGamma/dEdM}) with 
Gaussians in $M$ and $M_{3 \pi}$:
\begin{equation}
\int\,dM_{3\pi}' \,dM'\,
\frac{d\Gamma[J/\psi\, \omega]}{dM_{3\pi}'\,dM'}\,
\frac{e^{-(M-M')^2/2\sigma_X^2}}{\sqrt{2\pi}\sigma_X}
\frac{e^{-(M_{3\pi}-M'_{3\pi})^2/2\sigma_\omega^2}}{\sqrt{2\pi}\sigma_\omega}
\,.
\label{dGamma_smeared}
\end{equation}

Although it was not stated explicitly in Ref.~\cite{delAmoSanchez:2010jr},
the Babar data and histograms in Fig.~\ref{fig:Babar} are uncorrected 
for acceptances and efficiencies \cite{Dunwoody}.
This can be deduced from the fact that the central value of 
each data point for the combined distributions from $B^+$ and $B^0$ decay
in Fig.~4c of Ref.~\cite{delAmoSanchez:2010jr}
is equal to the sum of the central values of the data points for the separate
distributions from $B^+$ decay and $B^0$ decay in Figs.~4a and 4b.
Since the Babar data shown in Fig.~\ref{fig:Babar} are uncorrected,
direct comparisons with theoretical distributions for $M_{3\pi}$ 
are not appropriate.

In Ref.~\cite{delAmoSanchez:2010jr}, the generator for Babar's 
P-wave Monte Carlo differs from the generator for the S-wave Monte Carlo 
by a multiplicative factor of $q^2/(1 + R^2 q^2)$, where $q$
is the relative momentum of the $J/\psi$ and $R= 3~{\rm GeV}^{-1}$.
However the generator used for Babar's S-wave Monte Carlo 
is not stated in Ref.~\cite{delAmoSanchez:2010jr}.
Based on the limited information provided, 
a plausible guess is that the generator is equivalent to 
Eq.~(\ref{dGamma/dEdM}) with the resonance factor
$|\sum C_i f_{i2}|^2$ replaced by $\delta(M-M_X)$,
where $M_X = 3873.0$~MeV is the central value of the $X(3872)$ mass
from Babar's fit to the $J/\psi\, \pi^+ \pi^- \pi^0$ 
invariant mass distribution.
This central value corresponds to a negative binding energy
$E_X = -1.1$~MeV, but it is consistent within errors 
with the small positive binding energy in Eq.~(\ref{EX}).

\subsection{Previous theoretical analyses}

\begin{figure}[t]
\vspace*{-0.0cm}
\centerline{\includegraphics*[width=\columnwidth,angle=0,clip=true]{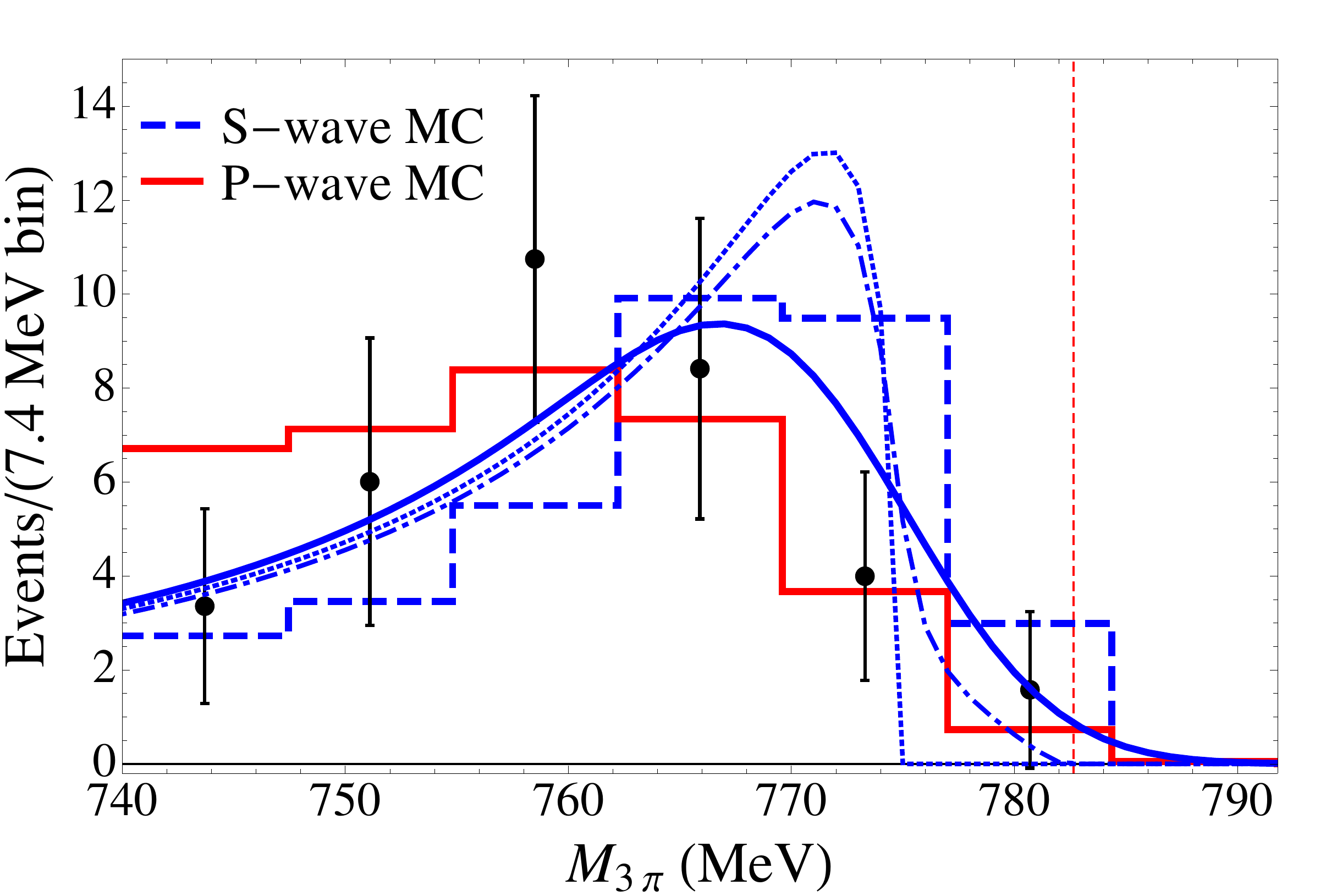}}
\vspace*{-0.0cm}
\caption{(Color online) 
Distribution of $M_{3\pi}$ integrated 
over the $J/\psi\, \pi^+\pi^-\pi^0$ 
invariant mass $M$ from 3862.5~MeV to 3882.5~MeV.
The data points and histograms are as described in Fig.~\ref{fig:Babar}.
The three curves are calculated from Eq.~(\ref{dGamma/dEdM}) 
using (1) the resonance factor $\delta(E)$ and no experimental resolution  
(dotted curve), as in Ref.~\cite{Hanhart:2011tn},
(2) a Breit-Wigner resonance factor with width 1.7~MeV 
and no experimental resolution (dash-dotted curve),
as in Ref.~\cite{Faccini:2012zv},
(3) the resonance factor $\delta(E)$ and experimental resolution of 
5.6~MeV in $M_{3 \pi}$ (solid curve).  
The vertical dashed line marks the position 
of the center of the $\omega$ resonance.
}
\label{fig:M3pi}
\end{figure}

There have been two previous theoretical analyses \cite{Hanhart:2011tn,Faccini:2012zv}
of the $M_{3\pi}$ distribution shown in Fig.~\ref{fig:Babar}.
Both analyses were based on the incorrect implicit
assumption that the Babar data were corrected for acceptances
and efficiencies.  Both analyses also ignored the experimental resolution.

Hanhart et al.\ carried out a combined analysis of data from the Belle
collaboration on the decay into $J/\psi\, \rho$ and from the Babar
collaboration on the decay into $J/\psi\, \omega$, comparing the
options $1^{++}$ and $2^{-+}$ for the quantum numbers of the $X(3872)$
\cite{Hanhart:2011tn}.  The Belle data was the distribution of $M_{2
  \pi}$ from $X \to J/\psi\, \pi^+\pi^-$, and it consisted of
approximately 200 events in 19 bins \cite{Choi:2011fc}.  The Babar
data was the distribution of $M_{3 \pi}$ from $X \to J/\psi\,
\pi^+\pi^- \pi^0$, and it consisted of only 25 events in 6 bins
\cite{delAmoSanchez:2010jr}.  The theoretical distributions for $M_{2
  \pi}$ and $M_{3 \pi}$ in Ref.~\cite{Hanhart:2011tn} take into
account $\rho-\omega$ mixing and the energy dependence of the $\rho$
and $\omega$ widths.  For the mass of the $X(3872)$, the authors used
$M_X = 3871.5$~MeV, and they ignored its width.  They also ignored the
experimental resolutions of the invariant masses $M_{2 \pi}$ and $M_{3
  \pi}$ of the pions and $M$ of the system consisting of $J/\psi$ and
pions.  The resulting distribution for $M_{3 \pi}$ drops to 0 sharply
at 775~MeV.  For the S-wave case, it is well-approximated by
Eq.~(\ref{dGamma/dEdM}) with the resonance factor $|\sum C_i
f_{i2}|^2$ replaced by $\delta(E)$.  This simple distribution is
illustrated in Fig.~\ref{fig:M3pi}, where it has been normalized so
that the area under the curve is 34 events.  The shape of the curve
does not resemble that of the Babar data or Babar's S-wave Monte Carlo
histogram, primarily because the experimental resolution  on $M_{3\pi}$
was ignored.  As a measure of the quality of the combined fit,
Ref.~\cite{Hanhart:2011tn} used the $\chi^2$ per degree of freedom for
the Babar and Belle data sets.  Given the large error bars in the
Babar data and the small number of data points, this measure is
sensitive only to the total number of $J/\psi\, \pi^+\pi^- \pi^0$
events and not to the shape of the $M_{3 \pi}$ distribution.  The
authors concluded from their analysis that the combined Belle and
Babar data favor the quantum numbers $1^{++}$.

Faccini et al.\ carried out an analysis \cite{Faccini:2012zv}
that also included the Belle data on angular distributions for the decay into
$ J/\psi\, \pi^+\pi^-$ \cite{Choi:2011fc}.
The authors used $M_X = 3872$~MeV,
and they took the width of the $X(3872)$ to be 1.7~MeV.
This is larger than the upper bound on the width in Eq.~(\ref{GammaX}).
They ignored the experimental resolutions of $M_{2 \pi}$, $M_{3 \pi}$, and 
$M$.  For the S-wave case, the resulting distribution for $M_{3 \pi}$ 
is well-approximated by Eq.~(\ref{dGamma/dEdM}) with 
$|\sum C_i f_{i2}|^2$ replaced by a Breit-Wigner function of $E$ 
centered at $E=0$ with width 1.7~MeV.
This distribution is illustrated in Fig.~\ref{fig:M3pi}, 
where it has been normalized so that the area under the curve is 34 events.
The sharp cutoff on $M_{3 \pi}$ at 775~MeV in Ref.~\cite{Hanhart:2011tn}
has been replaced by a tail 
from the Breit-Wigner that extends up to about 785~MeV.
The shape of the curve does not resemble that of the Babar data
or Babar's S-wave Monte Carlo  histogram,
primarily because the experimental resolution on $M_{3\pi}$ was ignored.
In the combined fit to the $J/\psi\, \pi^+\pi^-$ and $J/\psi\, \pi^+\pi^- \pi^0$ data,
the $\chi^2$ per degree of freedom favors the quantum numbers $1^{++}$,
but this measure is sensitive only to the total number of 
$J/\psi\, \pi^+\pi^- \pi^0$ events. According to Ref.~\cite{Faccini:2012zv},
the analysis of the $J/\psi\, \pi^+\pi^- \pi^0$ data alone excludes  $1^{++}$
at the 99.9\% confidence level.  
However this probability should not be taken seriously,
because experimental resolution was ignored in the analysis
and because the Babar data were not corrected 
for acceptances and efficiencies.

To take into account the resolution in the experiment 
of Ref.~\cite{delAmoSanchez:2010jr},
the distribution in Eq.~(\ref{dGamma/dEdM}) should be convoluted with 
a Gaussian in $M$ of width $\sigma_X=6.7$~MeV and a Gaussian in $M_{3 \pi}$ of width $\sigma_\omega=5.6$~MeV as in Eq.~(\ref{dGamma_smeared}).
The $M_{3 \pi}$ distribution is then obtained by integrating over
$M$ from 3862.5~MeV to 3882.5~MeV.
If we assume the resonance factor $|\sum C_i f_{i2}|^2$ is dominated by a region near the 
$D^{*0} \bar D^0$ threshold whose width is small compared to 6.7~MeV,
we can replace the resonance factor by a delta function at $E=0$.
The effect of integrating over the 20~MeV range of $E$
is then to constrain $M$ to be equal to the $D^{*0} \bar D^0$ threshold.
The experimental resolution on $M$ appears only in a multiplicative factor,
so it does not affect the shape of the $M_{3 \pi}$ distribution.
The resulting $M_{3 \pi}$ distribution is shown in Fig.~\ref{fig:M3pi}, 
where it has been normalized so that the area under the curve is 34 events.
The shape of this distribution is much closer to both
the Babar data and Babar's S-wave Monte Carlo than the distributions 
in which energy resolution was ignored.

Both of the previous theoretical analyses took into account the energy dependence 
of the width $\Gamma_\omega$ of the $\omega$ 
resonance \cite{Hanhart:2011tn,Faccini:2012zv}.  The energy dependence 
comes primarily from the total phase space for the decay
$\omega^* \to \pi^+ \pi^- \pi^0$.  
The phase space increases by about 2\% 
as the invariant mass increases by $\Gamma_\omega \approx 8.5$~MeV 
from $M_\omega \approx 783$~MeV to $M_\omega + \Gamma_\omega$.
Thus the effect of the energy-dependent width is not very dramatic.

\subsection{Resonance factor}

Since the Babar data in Fig.~\ref{fig:Babar} is uncorrected for
acceptances and efficiencies, direct comparisons with theoretical
$M_{3\pi}$ distributions are not appropriate.  However, given the
relatively low probability for Babar's S-wave Monte Carlo, it is worth
asking whether there are aspects of the $X(3872)$ resonance that could
improve the agreement between the S-wave Monte Carlo and the data.
Better agreement could have been obtained with a generator that gives
an $M_{3\pi}$ distribution whose peak is shifted lower by about 10~MeV
by suppressing the distribution above 770~MeV.
The differential rate for S-wave coupling to $J/\psi\, \omega$
in Eq.~(\ref{dGamma/dEdM}) implies that the $M_{3\pi}$ distribution has the form
\begin{eqnarray}
\frac{d\Gamma[J/\psi\, \omega]}{dM_{3\pi}}= 
\frac{\Gamma_\omega}
{2 \pi [(M_{3\pi} - M_\omega)^2 + \Gamma_\omega^2/4]}
\int_{E_{\rm min}} \!\!\!\!\!\!dE~
\big| \mbox{$\sum$}_{i} C_i f_{i2}(E) \big|^2 q(E) ,
\label{dGamma/dM}
\end{eqnarray}
where $q(E) = [2 \mu_{\psi \omega} 
(E - \delta_{\psi \omega} - M_{3\pi} + M_\omega)]^{1/2}$,
$E_{\rm min} = M_{3\pi} - M_\omega + \delta_{\psi \omega}$,
and the upper endpoint of
the integral over $E$ is well above the $D^* \bar D$ threshold region.
The last factor in Eq.~(\ref{dGamma/dM}) is the integral of the line shape
weighted by the relative momentum $q$.
We wish to determine whether the dependence of this factor on $M_{3\pi}$
could improve the agreement between the S-wave Monte Carlo
and the Babar data in Fig.~\ref{fig:Babar}.
The generator for Babar's P-wave Monte Carlo produced a downward 
shift in the peak of the $M_{3\pi}$ distribution by about 10~MeV
through an additional multiplicative factor of $q^2/(1 + R^2 q^2)$. 
If the line shape can be approximated by a delta function near $E=0$,
the factor of $q^2$ has a zero at  
$M_{3\pi} = M_\omega - \delta_{\psi \omega}  \approx 775$~MeV.
We wish to determine whether a comparable shift can be 
produced instead by a change in the resonance factor.

\subsubsection{Universal resonance factor}

\begin{figure}[t]
\vspace*{-0.0cm}
\centerline{\includegraphics*[width=\columnwidth,angle=0,clip=true]{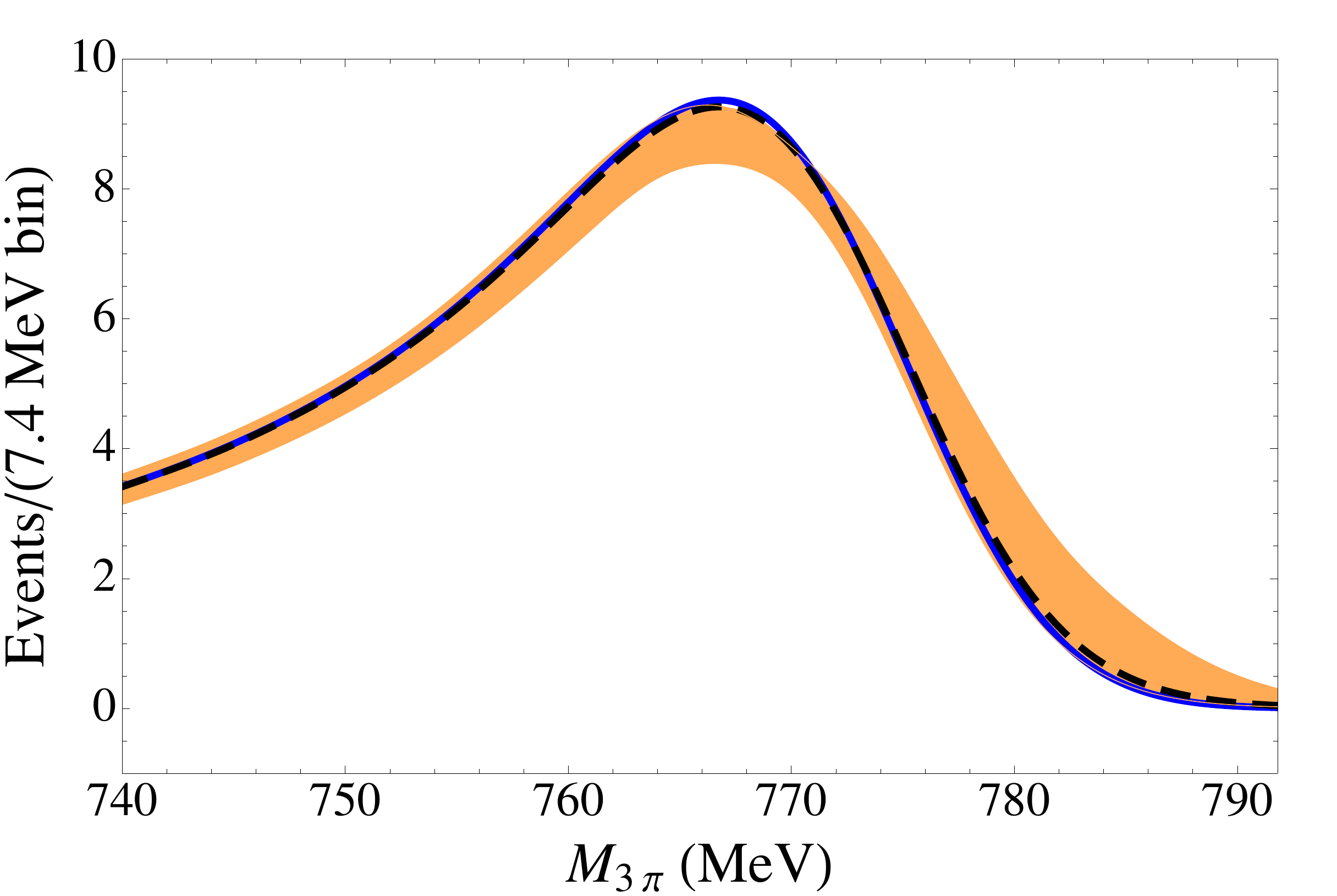}}
\vspace*{-0.0cm}
\caption{(Color online) Distribution of $M_{3\pi}$ integrated over the
  $J/\psi\, \pi^+\pi^-\pi^0$ invariant mass $M$ from 3862.5~MeV to 3882.5~MeV.  
  The experimental resolutions of 6.7~MeV in $M$ and 5.6~MeV in $M_{3 \pi}$ 
  are taken into account.  The curves are calculated using 
  the binding energy  $E_X=0.26$~MeV and  a resonance factor that is 
either a delta function at $E=-E_X$ (blue solid curve)
or the universal resonance factor with the minimal width $\Gamma_X=0.066$~MeV 
(black dashed curves).  The shaded band takes into account
 variations in $E_X$ from 0 to 0.65~MeV and in $\Gamma_X$ from 0.066 to 1.2~MeV.  }
\label{fig:M3piB}
\end{figure}

According to Eq.~(\ref{sumCf-gamma}), the resonance factor sufficiently near the  
$D^{*0} \bar D^0$ threshold has the universal line shape
$|-\gamma +\kappa(E)|^{-2}$.  This universal resonance factor
can not be approximated by a delta 
function in $E$, because it has power-law tails that decrease as 
$1/|2 \mu_0 E|$ at large $|E|$.
When integrated over a smooth distribution in $E$, it
can be approximated by the sum of a Lorentzian in $E$ and $1/|E|$ tails
at positive and negative energies.  The Lorentzian is centered at $-E_X$ 
with  width $\Gamma_X$, where $E_X$ and $\Gamma_X$ are the binding energy 
and width of the $X(3872)$, and it can be approximated by 
$(4 \pi E_X/\mu_0 \Gamma_X) \delta(E+ E_X)$.  
The tails have the form $1/|2 \mu_0 E|$ that extend from small $|E|$
of order $E_X$ to large $|E|$ of order $\delta_1$ or $\delta_{\psi \omega}$.  
The relative sizes of the integrals of the line shape over the peak
near $-E_X$ and over the tails are approximately $4 \pi E_X/\Gamma_X$ 
and $\ln(\delta_1/E_X)$, so the contribution from the tails can be significant.
The $M_{3 \pi}$ distributions for the universal resonance factor 
$|-\gamma +\kappa(E)|^{-2}$ and for $\delta(E+ E_X)$, 
with the resolutions in $M$ and $M_{3 \pi}$ taken into account,  
are compared in Fig.~\ref{fig:M3piB}.
Replacing $\delta(E+ E_X)$ by the universal resonance factor 
gives a negligible shift in the position of the peak in $M_{3 \pi}$.  
The variations in the $M_{3 \pi}$ distribution for the universal resonance factor 
from the uncertainties in $E_X$ and $\Gamma_X$ are also shown in
 Fig.~\ref{fig:M3piB}.  The small shifts in the position of the peak come primarily
 from the variations in $E_X$.

\subsubsection{Interference between scattering channels}

 Babar's P-wave Monte Carlo produced a significant shift in the 
position of the peak of the $M_{3\pi}$ distribution
through a multiplicative factor of $q^2$, 
which has a zero at 
$E = \delta_{\psi \omega} + M_{3\pi} - M_\omega$.
It is plausible that a significant shift could also be 
produced by an approximate zero in the resonance factor instead.
An approximate zero of $\sum C_i f_{i2}(E)$ could arise 
from interference between the scattering channels.
To suppress the region of $M_{3\pi}$ above 765~MeV,
the approximate zero of $\sum C_i f_{i2}(E)$ would have to be
at an energy $E_0$ above the $D^{*0} \bar D^0$ threshold.
The shape of the general resonance factor $|\sum C_i f_{i2}|^2$
depends on the ratios  $C_1/C_0$ and $C_2/C_0$ of the
complex short-distance factors and on the scattering parameters 
$\gamma_0$, $\gamma_1$, $\gamma_X$, and $\gamma_V$.  
The short-distance factors $C_i$ can be complex, so an 
approximate zero requires a fine tuning of these coefficients.

We can examine this possibility with a simplified form
of the resonance factor that has fewer adjustable parameters. 
We set $C_2=0$, which implies that production of the 
$X(3872)$ resonance is dominated by the creation of charm meson pairs 
at short distance rather than $J/\psi\,\omega$.
This assumption  is motivated by the suppression
of $C_2$ suggested by the estimates in Section~\ref{sec:lineshapeSD}.
The resonance factor then reduces to $C_0 f_{02} + C_1 f_{12}$,
where the scattering amplitudes $f_{02}(E)$ and $f_{12}(E)$ 
are given in Eqs.~(\ref{f02}) and (\ref{f12}).
The numerator of the resonance factor is proportional to
$C_0 (-\gamma_1+\kappa_1) - C_1(-\gamma_1+\kappa)$. 
If there is an interference zero at $E_0$,
the short-distance coefficients must satisfy
\begin{equation}
\frac{C_1}{C_0} = \frac{-\gamma_1+\kappa_1 (E_0)}{-\gamma_1+\kappa(E_0)}.
\label{C1overC0}
\end{equation}
We further assume $|\gamma_V| \gg |\kappa_{\psi\omega}(E)|$, 
which implies that the $J/\psi\,\omega$ channel would not have a bound state
near threshold in the absence of the $X(3872)$ resonance.
The denominator $D(E)$ in $f_{02}(E)$ and $f_{12}(E)$ then
reduces to the denominator $D'(E)$ for the 2-channel case
in Eq.~(\ref{DenBL}), with $\gamma_0$ replaced by
\begin{equation}
\gamma_0^{\text{eff}} =  \gamma_0- 2\gamma_X^2/\gamma_V.
\label{gamma0eff}
\end{equation}
The scattering parameter $\gamma_1$ is determined by the 
value of $\gamma$ as in Eq.~(\ref{gamma}), which reduces to
\begin{equation}
\gamma =
\frac{ 2 \gamma_0^{\text{eff}} \gamma_1 
- (\gamma_0^{\text{eff}} +\gamma_1) \kappa_1(0)}
{\gamma_0^{\text{eff}} + \gamma_1 - 2 \kappa_1(0)}.
\label{gamma-2ch}
\end{equation}
Thus $\gamma_0^{\text{eff}}$ is the only scattering
parameter that affects the shape of the resonance factor.

The adjustable parameters in our simplified resonance factor are
$\gamma_0^{\text{eff}}$ and $E_0$.  We wish to determine whether the
peak in the $M_{3\pi}$ distribution can be shifted downward by about
10~MeV by adjusting these parameters.  We therefore consider $E_0$ in
the range between 0 and 10~MeV. For some values of
$\gamma_0^{\text{eff}}$, there is another resonance with energy
below $X(3872)$, which is inconsistent with observations.  Demanding
no such resonance constrains $\gamma_0^\text{eff}<\kappa_1(0)/2\approx
60$~MeV.  There can also be a substantial enhancement near
$E=\delta_1$ from a cusp associated with the opening up of the $D^{*+}
D^-$ threshold.  This has the opposite effect of the suppression above
the threshold that we want.  To avoid the enhancement, we require
$|\gamma_0^\text{eff}+\gamma_1| > 2|\kappa_1(0)|$, which combined with
Eq.~\eqref{gamma-2ch} implies $\gamma_0^\text{eff}> 50$~MeV or
$\gamma_0^\text{eff}< -300$~MeV.  For the given regions of $E_0$ and
$\gamma_0^\text{eff}$, we are unable to obtain a significant shift in
the peak of the $M_{3\pi}$ distribution to lower mass.

\subsubsection{Interference from tail of $\chi_{c1}(2P)$ resonance}

 Another way to suppress the resonance factor for $E$ above the $D^{*0}
\bar D^0$ threshold is through interference with the low-energy tail
of the $\chi_{c1}(2P)$ resonance.  The inclusive line shape 
in the isospin-0
channel is given in Eq.~\eqref{dGamma-psiomegachi}. 
It can be resolved into contributions proportional to the imaginary parts of 
$\gamma_0$, $-\nu$, and $g$ by using Eq.~\eqref{Imgammachi}.
The line shape in the $J/\psi\, \omega$ channel can be obtained
by replacing the imaginary parts of $\gamma_0$, $-\nu$, and $g$
in Eq.~(\ref{Imgammachi}) by the contributions 
to those imaginary parts from the $J/\psi\, \omega$ channel,
and allowing those imaginary parts to be energy-dependent.
For example, the substitution for Im$(\gamma_0)$
is given in Eq.~(\ref{Imgamma0-psiomega}).
If we use the assumption $|\gamma_V| \gg |\kappa_{\psi \omega}(E)|$,
the substitution for Im$(\gamma_0)$ reduces to
$- {\rm Im} \kappa_{\psi \omega}(E)$ multiplied by a constant. 
The substitutions for  $-\nu$ and $g$ would reduce to similar forms.
The expression in Eq.~\eqref{Imgammachi} then reduces to 
the product of the resonance factor $1/ |E-\nu+g^2 \gamma_0|^2$,
the threshold factor $- {\rm Im} \kappa_{\psi \omega}(E)$,
and a quadratic function of the energy $E$.
The maximum possible interference effect corresponds 
to total destructive interference at some energy 
$E_0$ above the $D^{*0} \bar D^0$ threshold.
In this case, the quadratic function of $E$ reduces to $(E-E_0)^2$ 
and the expression in Eq.~(\ref{Imgammachi})
reduces to
\begin{equation}
\left[ \text{Im}\left(\frac{1}{\gamma_0}+\frac{g^2}{E-\nu} \right)^{-1}
\right]_{J/\psi\, \omega} \longrightarrow\, 
\frac{ -\text{Im}\,\kappa_{\psi \omega} (E)}
    { |E-\nu+g^2 \gamma_0|^2}
\,\frac{2\gamma_X^2}{|\gamma_V|^2}
\, (E - E_0)^2 ,
\label{Imgamma0-last}
\end{equation}
We vary the position of interference zero by changing $E_0$ 
between 0 and 10~MeV. 
We are unable to obtain a significant shift in the peak 
of the distributions to smaller $M_{3\pi}$ by tuning the 
interference position $E_0$.
We are also unable to obtain a significant shift using zeroes 
in both the $X(3872)$ resonance factor $| \sum C_i f_{i2}|^2$
and the $\chi_{c1}(2P)$ resonance factor in Eq.~\eqref{Imgamma0-last}. 

\section{Summary}
\label{sec:summary}

The quantum numbers of the $X(3872)$ have been definitely 
established as $1^{++}$ from analyses of the $J/\psi\, \pi^+\pi^-$ decay channel.  
This settles an issue raised by a Babar analysis of the $M_{3\pi}$ distribution 
for the $J/\psi\, \pi^+\pi^- \pi^0$ decay channel that preferred 
$2^{-+}$ over $1^{++}$ \cite{delAmoSanchez:2010jr}.
We pointed out that in the Babar analysis, the quoted values of
$\chi^2$ were not minimized with respect to the adjustable normalizations
of the Monte Carlo distributions.  Upon minimization of the $\chi^2$, 
the probability for $1^{++}$ is increased significantly 
from 7.1\% to 18.7\% while the probability for $2^{-+}$ is
increased only slightly from  61.9\% to 66.2\%.  
Thus the preference for $2^{-+}$ over $1^{++}$ 
was overstated in Ref.~\cite{delAmoSanchez:2010jr}.

For the benefit of future analyses of the $X(3872)$ resonance in
the $J/\psi\, \pi^+\pi^-\pi^0$ decay channel, we considered
whether a more accurate description of the resonance could 
have further improved the agreement between the Babar data 
and the Babar S-wave Monte Carlo for the $1^{++}$ case.
To describe the resonance more accurately, 
we derived the low-energy scattering amplitudes due to 
S-wave couplings between the three channels in Eq.~(\ref{ch012}): 
neutral  charm meson pairs, charged charm meson pairs, 
and $J/\psi\,\omega$.   
We also considered how the scattering would be affected
by an additional 
$\chi_{c1}(2P)$ resonance with quantum numbers $1^{++}$.
We used the scattering amplitudes to derive the line shape
for the $J/\psi\, \pi^+\pi^- \pi^0$ decay channel
and also the $M_{3\pi}$ distribution.

The Babar P-wave Monte Carlo that was  preferred by the Babar data
over the S-wave Monte Carlo gave an  $M_{3\pi}$ distribution 
whose peak was about 10~MeV lower.
We considered several mechanisms for shifting the peak 
for the S-wave case to lower values of $M_{3\pi}$.
We considered the effects of the power-law tails of the
universal scattering amplitude.  We considered interference between
the charm-meson  scattering channels. We also considered
the interference from the tail of a higher $\chi_{c1}(2P)$ resonance. 
For all these mechanisms, the $M_{3\pi}$ distribution was robust 
against a shift in the peak shift to lower values.
We conclude that, given the resolution in the Babar 
experiment, a more accurate description 
of the $X(3872)$ resonance in the $J/\psi\, \pi^+\pi^-\pi^0$ decay channel
is not essential.
The effects we have considered may however be important 
in future analyses of this decay channel with higher resolution.

\begin{acknowledgments}
This research was supported in part by the U.S.\ Department of Energy
under grant DE-FG02-91-ER40690 and under Contract DE-FG02-94ER40818.
We thank A.~Mokhtar and W.~Dunwoody for providing the Babar results
in Fig.~\ref{fig:Babar}.
\end{acknowledgments}




\begin{thebibliography}{}


\bibitem{Choi:2003ue}
S.~K.~Choi {\it et al.}  [Belle Collaboration],
Phys.\ Rev.\ Lett.\  {\bf 91}, 262001 (2003)
 [arXiv:hep-ex/0309032].

\bibitem{Close:2003sg} 
  F.~E.~Close and P.~R.~Page,
  Phys.\ Lett.\ B {\bf 578}, 119 (2004)
  [hep-ph/0309253].
  
\bibitem{Voloshin:2003nt} 
  M.~B.~Voloshin,
  Phys.\ Lett.\ B {\bf 579}, 316 (2004)
  [hep-ph/0309307].
  
\bibitem{Abe:2005ix}
  K.~Abe {\it et al.} [Belle Collaboration],
  arXiv:hep-ex/0505037.

\bibitem{Aubert:2006aj}
  B.~Aubert {\it et al.}  [BABAR Collaboration],
  Phys.\ Rev.\ D {\bf 74} (2006) 071101
  [hep-ex/0607050].

\bibitem{Abulencia:2006ma}
  A.~Abulencia {\it et al.}  [CDF Collaboration],
  Phys.\ Rev.\ Lett.\  {\bf 98} (2007) 132002
  [hep-ex/0612053].

\bibitem{Aaij:2013zoa} 
  R.~Aaij {\it et al.}  [LHCb Collaboration],
  arXiv:1302.6269.
  
\bibitem{Barnes:2003vb} 
  T.~Barnes and S.~Godfrey,
  Phys.\ Rev.\ D {\bf 69}, 054008 (2004)
  [hep-ph/0311162].
  
\bibitem{Jia:2010jn} 
  Y.~Jia, W.~-L.~Sang and J.~Xu,
  arXiv:1007.4541.
  
\bibitem{Burns:2010qq} 
  T.~J.~Burns, F.~Piccinini, A.~D.~Polosa and C.~Sabelli,
  Phys.\ Rev.\ D {\bf 82}, 074003 (2010)
  [arXiv:1008.0018].

\bibitem{Kalashnikova:2010hv} 
  Y.~.S.~Kalashnikova and A.~V.~Nefediev,
  Phys.\ Rev.\ D {\bf 82}, 097502 (2010)
  [arXiv:1008.2895].
  
\bibitem{Hanhart:2011tn} 
  C.~Hanhart, Y.~.S.~Kalashnikova, A.~E.~Kudryavtsev and A.~V.~Nefediev,
  Phys.\ Rev.\ D {\bf 85}, 011501 (2012)
  [arXiv:1111.6241].

\bibitem{delAmoSanchez:2010jr} 
  P.~del Amo Sanchez {\it et al.}  [BABAR Collaboration],
  Phys.\ Rev.\ D {\bf 82}, 011101 (2010)
  [arXiv:1005.5190].

\bibitem{Stapleton:2009ey} 
  E.~Braaten and J.~Stapleton,
  Phys.\ Rev.\ D {\bf 81}, 014019 (2010)
  [arXiv:0907.3167].

\bibitem{Aaltonen:2009vj} 
  T.~Aaltonen {\it et al.}  [CDF Collaboration],
  Phys.\ Rev.\ Lett.\  {\bf 103}, 152001 (2009)
  [arXiv:0906.5218]. 
  
\bibitem{Choi:2011fc} 
S.~-K.~Choi  {\it et al.},
  Phys.\ Rev.\ D {\bf 84}, 052004 (2011)
  [arXiv:1107.0163].

\bibitem{Aaij:2011sn} 
  R.~Aaij {\it et al.}  [LHCb Collaboration],
  Eur.\ Phys.\ J.\ C {\bf 72}, 1972 (2012)
  [arXiv:1112.5310].
  
\bibitem{Aubert:2008gu} 
  B.~Aubert {\it et al.}  [BABAR Collaboration],
  Phys.\ Rev.\ D {\bf 77}, 111101 (2008)
  [arXiv:0803.2838].
  
\bibitem{Beringer:1900zz} 
  J.~Beringer {\it et al.}  [Particle Data Group Collaboration],
  Phys.\ Rev.\ D {\bf 86}, 010001 (2012).
  
\bibitem{Aaij:2013uaa} 
  R.~Aaij {\it et al.}  [LHCb Collaboration],
  arXiv:1304.6865.

\bibitem{Tomaradze:2012iz} 
  A.~Tomaradze, S.~Dobbs, T.~Xiao, K.~K.~Seth and G.~Bonvicini,
  arXiv:1212.4191. 

\bibitem{Braaten:2007dw} 
  E.~Braaten and M.~Lu,
  Phys.\ Rev.\ D {\bf 76}, 094028 (2007)
  [arXiv:0709.2697].

\bibitem{Braaten:2007ft} 
  E.~Braaten and M.~Lu,
  Phys.\ Rev.\ D {\bf 77}, 014029 (2008)
  [arXiv:0710.5482].
   
  
\bibitem{Braaten:2004rn} 
  E.~Braaten and H.~-W.~Hammer,
  Phys.\ Rept.\  {\bf 428}, 259 (2006)
  [cond-mat/0410417].

\bibitem{Aubert:2007vj} 
  B.~Aubert {\it et al.}  [BaBar Collaboration],
  Phys.\ Rev.\ Lett.\  {\bf 101}, 082001 (2008)
  [arXiv:0711.2047].
  
\bibitem{Uehara:2009tx} 
  S.~Uehara {\it et al.}  [Belle Collaboration],
  Phys.\ Rev.\ Lett.\  {\bf 104}, 092001 (2010)
  [arXiv:0912.4451].
  
\bibitem{Lees:2012xs} 
  J.~P.~Lees {\it et al.}  [BABAR Collaboration],
  Phys.\ Rev.\ D {\bf 86}, 072002 (2012)
  [arXiv:1207.2651].

\bibitem{Artoisenet:2010va} 
  P.~Artoisenet, E.~Braaten and D.~Kang,
  Phys.\ Rev.\ D {\bf 82}, 014013 (2010)
  [arXiv:1005.2167].
    
\bibitem{Hanhart:2011jz} 
  C.~Hanhart, Y.~.S.~Kalashnikova and A.~V.~Nefediev,
  Eur.\ Phys.\ J.\ A {\bf 47}, 101 (2011)
  [arXiv:1106.1185].

\bibitem{Braaten:2005jj}
  E.~Braaten and M.~Kusunoki,
  Phys.\ Rev.\ D {\bf 72}, 014012 (2005)
  [arXiv:hep-ph/0506087].

\bibitem{delAmoSanchez:2010pg} 
  P.~del Amo Sanchez {\it et al.}  [BABAR Collaboration],
  Phys.\ Rev.\ D {\bf 83}, 032004 (2011)
  [arXiv:1011.3929].

\bibitem{Dunwoody}
W.~Dunwoody, private communication.

\bibitem{Aubert:2006zb} 
  B.~Aubert {\it et al.}  [BABAR Collaboration],
  Phys.\ Rev.\ D {\bf 74}, 012001 (2006)
  [hep-ex/0604009].

\bibitem{Faccini:2012zv} 
  R.~Faccini, F.~Piccinini, A.~Pilloni and A.~D.~Polosa,
  Phys.\ Rev.\ D {\bf 86}, 054012 (2012)
  [arXiv:1204.1223].
   
\end{thebibliography}
\end{document}